\documentstyle[12pt]{article}

\newcommand{\sect}[1]{\setcounter{equation}{0}\section{#1}}

\def\bseq{\begin{subequation}}  
\def\eseq{\end{subequation}}
\def\bsea{\begin{subeqnarray}}  
\def\esea{\end{subeqnarray}}

\def\Hat#1{\widehat{#1}}                        
\def\Bar#1{\overline{#1}}                       
\def\lin{\vrule width0.5pt height5pt depth1pt}
\def\dpx{{{ =\hskip-3.75pt{\lin}}\hskip3.75pt }}
\def\dvm{\raisebox{-.45ex}{\rlap{$=$}} }
\def\DM{{\scriptsize{\dvm}}~~}
\def\Tilde#1{\widetilde{#1}}                    

\newcommand{\beq}{\begin{equation}}
\newcommand{\eeq}{\end{equation}}
\newcommand{\bea}{\begin{eqnarray}}
\newcommand{\eea}{\end{eqnarray}}
\newcommand{\ena}{\end{eqnarray}}

\newcommand {\non}{\nonumber}

\renewcommand{\a}{\alpha}
\renewcommand{\b}{\beta}

\renewcommand{\d}{\delta}
\newcommand{\th}{\theta}
\newcommand{\q}{\theta}

\newcommand{\pa}{\partial}
\newcommand{\di}{\partial}
\newcommand{\g}{\gamma}
\newcommand{\G}{\Gamma}

\newcommand{\D}{\Delta}
\newcommand{\e}{\epsilon}

\renewcommand{\l}{\lambda}

\newcommand{\m}{\mu}

\newcommand{\n}{\nu}

\newcommand{\f}{\phi}
\newcommand{\vf}{\varphi}
\newcommand{\F}{\Phi}

\newcommand{\vp}{\varpi}

\newcommand{\s}{\sigma}
\renewcommand{\S}{\Sigma}

\renewcommand{\o}{\omega}
\renewcommand{\O}{\Omega}

\newcommand{\Db}{\bar{D}}

\newcommand{\Fb}{\bar{F}}

\newcommand{\Pb}{\bar{P}}
\newcommand{\phib}{\bar{\phi}}

\newcommand{\chib}{\bar{\chi}}

\newcommand{\calP}{{\cal P}}

\newcommand{\calX}{{\cal X}}
\newcommand{\calPb}{\bar{{\cal P}}}

\newcommand{\Phib}{\bar{\Phi}}

\def\Mb{\kern 2pt\mathchoice
            {
             \vbox{\hrule width10pt height 0.4pt depth 0pt
                 \kern 1.2pt\hbox{\kern -2pt$\displaystyle M$}}}
            {
                 \vbox{\hrule width10pt height 0.4pt depth 0pt
                 \kern 1.2pt\hbox{\kern -2pt$\textstyle M$}}}
            {
\vbox{\hrule width6pt height 0.4pt depth 0pt
                 \kern 1.0pt\hbox{\kern -2pt$\scriptstyle M$}}}
            {
                 \vbox{\hrule width5pt height 0.4pt depth 0pt
                 \kern 0.8pt\hbox{\kern -2pt$\scriptscriptstyle M$}}}}

\def\Sb{\kern 2pt\mathchoice
            {
                 \vbox{\hrule width6pt height 0.4pt depth 0pt
                 \kern 1.2pt\hbox{\kern -2pt$\displaystyle S$}}}
            {
                 \vbox{\hrule width6pt height 0.4pt depth 0pt
                 \kern 1.2pt\hbox{\kern -2pt$\textstyle S$}}}
            {
                 \vbox{\hrule width3.5pt height 0.4pt depth 0pt
                 \kern 1.0pt\hbox{\kern -2pt$\scriptstyle S$}}}
            {
                 \vbox{\hrule width3pt height 0.4pt depth 0pt
                 \kern 0.8pt\hbox{\kern -2pt$\scriptscriptstyle S$}}}}

\def\Rb{\kern 2pt\mathchoice
            {
                 \vbox{\hrule width5.5pt height 0.4pt depth 0pt
                 \kern 1.2pt\hbox{\kern -2.5pt$\displaystyle R$}}}
            {
                 \vbox{\hrule width5.5pt height 0.4pt depth 0pt
                 \kern 1.2pt\hbox{\kern -2.5pt$\textstyle R$}}}
            {
                 \vbox{\hrule width3.5pt height 0.4pt depth 0pt
                 \kern 1.0pt\hbox{\kern -2.2pt$\scriptstyle R$}}}
            {
                 \vbox{\hrule width3pt height 0.4pt depth 0pt
                 \kern 0.8pt\hbox{\kern -2.2pt$\scriptscriptstyle R$}}}}

  \def\pp{{\mathchoice
          {
              \kern 1pt%
              \raise 1pt
              \vbox{\hrule width5pt height0.4pt depth0pt
                    \kern -2pt
                    \hbox{\kern 2.3pt
                          \vrule width0.4pt height6pt depth0pt
                          }
                    \kern -2pt
                    \hrule width5pt height0.4pt depth0pt}%
                    \kern 1pt
           }
            {
              \kern 1pt%
              \raise 1pt
              \vbox{\hrule width4.3pt height0.4pt depth0pt
                    \kern -1.8pt
                    \hbox{\kern 1.95pt
                          \vrule width0.4pt height5.4pt depth0pt
                          }
                    \kern -1.8pt
                    \hrule width4.3pt height0.4pt depth0pt}%
                    \kern 1pt
            }
            {
              \kern 0.5pt%
              \raise 1pt
              \vbox{\hrule width4.0pt height0.3pt depth0pt
                    \kern -1.9pt  
                    \hbox{\kern 1.85pt
                          \vrule width0.3pt height5.7pt depth0pt
                          }
                    \kern -1.9pt
                    \hrule width4.0pt height0.3pt depth0pt}%
                    \kern 0.5pt
            }
            {
              \kern 0.5pt%
              \raise 1pt
              \vbox{\hrule width3.6pt height0.3pt depth0pt
                    \kern -1.5pt
                    \hbox{\kern 1.65pt
                          \vrule width0.3pt height4.5pt depth0pt
                          }
                    \kern -1.5pt
                    \hrule width3.6pt height0.3pt depth0pt}%
                    \kern 0.5pt
            }
        }}

  \def\mm{{\mathchoice
                       {
                             \kern 1pt
               \raise 1pt    \vbox{\hrule width5pt height0.4pt depth0pt
                                  \kern 2pt
                                  \hrule width5pt height0.4pt depth0pt}
                             \kern 1pt}
                       {
                            \kern 1pt
               \raise 1pt \vbox{\hrule width4.3pt height0.4pt depth0pt
                                  \kern 1.8pt
                                  \hrule width4.3pt height0.4pt depth0pt}
                             \kern 1pt}
                       {
                            \kern 0.5pt
               \raise 1pt
                            \vbox{\hrule width4.0pt height0.3pt depth0pt
                                  \kern 1.9pt
                                  \hrule width4.0pt height0.3pt depth0pt}
                            \kern 1pt}
                       {
                           \kern 0.5pt
             \raise 1pt  \vbox{\hrule width3.6pt height0.3pt depth0pt
                                  \kern 1.5pt
                                  \hrule width3.6pt height0.3pt depth0pt}
                           \kern 0.5pt}
                       }}

\def\pd{{\kern0.5pt
                   + \kern-5.05pt \raise5.8pt\hbox{$\textstyle.$}\kern
0.5pt}}

\def\pmd{{\kern0.5pt
                  \pm \kern-5.05pt \raise6.3pt\hbox{$\textstyle.$}\kern1.5pt}}

\def\md{{\mathchoice
   {
      {{\kern 1pt - \kern-6.2pt \raise5pt\hbox{$\textstyle.$}\kern 1pt}}}
    {
      {{\kern 1pt - \kern-6.2pt \raise5pt\hbox{$\textstyle.$}\kern 1pt}}}
    {
      {\kern0.5pt - \kern-5.05pt \raise3.4pt\hbox{$\textstyle.$}\kern0.5pt}}
    {
      {\kern0.5pt - \kern-5.05pt \raise3.4pt\hbox{$\textstyle.$}\kern0.5pt}}}}

\newcommand{\Apm}{A_+^{~-}}
\newcommand{\Amp}{A_-^{~+}}
\newcommand{\Apmd}{A_{\pd}^{~\md}}
\newcommand{\Ampd}{A_{\md}^{~\pd}}

\newcommand{\Ch}{\hat{C}}

\newcommand{\Cmpp}{\Ch_{-\pd}^{~~~\pp}}
\newcommand{\Cpmp}{\Ch_{+\md}^{~~~\pp}}
\newcommand{\Cpmm}{\Ch_{+\md}^{~~~\mm}}
\newcommand{\Cmpm}{\Ch_{-\pd}^{~~~\mm}}

\newcommand{\BAH}{\buildrel \leftarrow \over H}

\newcommand{\Dpparl}{\buildrel \leftrightarrow \over {\cal D}_{\pp}}
\newcommand{\Dmmarl}{\buildrel \leftrightarrow \over {\cal D}_{\mm}}

\newcommand{\ad}{{\dot{\alpha}}}
\newcommand{\bd}{{\dot{\beta}}}
\newcommand{\Del}{\nabla}
\newcommand{\Delb}{\bar{\nabla}}
\newcommand{\Delp}{\nabla_{+}}
\newcommand{\Delm}{\nabla_{-}}
\newcommand{\Delpd}{\nabla_{\pd}}
\newcommand{\Delmd}{\nabla_{\md}}
\newcommand{\Delpp}{\nabla_{\pp}}
\newcommand{\Delmm}{\nabla_{\mm}}
\newcommand{\Delsq}{\nabla^2}
\newcommand{\Delbsq}{{\bar{\nabla}}^2}
\newcommand{\Dela}{\nabla_{\alpha}}
\newcommand{\Delad}{\nabla_{\dot{\alpha}}}
\newcommand{\DelA}{\nabla_A}

\newcommand{\bD}{{\bf D}}

\newcommand{\bDpp}{{\bf D}_{\pp}}
\newcommand{\bDmm}{{\bf D}_{\mm}}

\newcommand{\calD}{{\cal D}}

\newcommand{\calDpp}{{\cal D}_\pp}
\newcommand{\calDmm}{{\cal D}_\mm}

\def\Sc{\scriptstyle}

\newcommand{\reff}[1]{(\ref{#1})}

\newcommand{\shalf}{{\frac{1}{2}}}
\newcommand{\sihalf}{{\frac{i}{2}}}
\newcommand{\half}{\frac{1}{2}}
\newcommand{\ihalf}{\frac{i}{2}}
\newcommand{\squart}{{\Sc\frac{1}{4}}}

\renewcommand{\thefootnote}{\fnsymbol{footnote}}

\begin{document}

\newpage
\begin{titlepage}
\begin{flushright}
{hep-th/9509021}\\
{BRX-TH-379}\\
{UMDEPP 96-21}
\end{flushright}
\vspace{2cm}
\begin{center}

{\large A STUDY OF GENERAL 2D, N=2 MATTER COUPLED TO
SUPERGRAVITY  IN SUPERSPACE }
\footnote{Research supported in part
by NSF grant \# PHY-93-41926 and \# PHY-92-22318}
${}$ \footnote {Supported in part by NATO Grant CRG-93-0789}
\\[.2in]
S. James Gates, Jr.\footnote{GATES@UMDHEP.UMD.EDU}   \\[.1in]
{\it Department of Physics\\
University of Maryland at College Park\\
College Park, MD 20742-4111, USA}\\[.15in]
{\rm {and}}\\[.15in]
M.~T.~Grisaru\footnote{GRISARU@BINAH.CC.BRANDEIS.EDU}\\[.04in]
{\rm {and}}\\[.04in]
M.~E.~Wehlau\footnote{MARCIA@ASTRO.QUEENSU.CA, \newline
\hbox to \hsize{Current address:
Mars Scientific Consulting, 28 Limeridge Dr.,
 Kingston,  ON CANADA K7K~6M3}}\\[.1in]
{\it  Physics Department\\
Brandeis University\\
Waltham, MA~~02254, USA}\\[1in]

{ABSTRACT}\\[.1in]
\end{center}
\begin{quote}
We study  two-dimensional  $N = 2$ supersymmetric
actions describing general  models of scalar and vector multiplets
coupled to supergravity.
\end{quote}

\vfill

\begin{flushleft}
September 1995

\end{flushleft}
\end{titlepage}

\newpage

\renewcommand{\thefootnote}{\arabic{footnote}}
\setcounter{footnote}{0}
\newpage
\pagenumbering{arabic}

\sect{Introduction}

Aspects of $2D$ local supersymmetry theories have been studied  most
extensively
in connection with realizations of local superconformal symmetry
and its applications to string theory \cite{A}, conformal field
theory \cite{B}, etc.  But even beyond these applications where superconformal
symmetry is of paramount importance, local supersymmetric actions
have  been observed to play an important role
for integrable systems \cite{C}.   On one hand, we have
 the observation \cite{OV} that $2D$,  $N = 2$ superstrings possess very close
 relations to self-dual target space theories.
On the other hand there is the well-known conjecture due to Atiyah that
suggests a relation between self-dual Yang-Mills gauge theory and integrable
models in less than four dimensions \cite{AWH}.  Some evidence  has been
presented
\cite{GN1}
to indicate that the supersymmetric extension of this suggestion is also
valid. It is in this spirit that one might believe
that a general study of $2D$ locally supersymmetric Lagrangian field theories
is worthwhile.

With this as our main motivation, we propose to answer the question, ``What
is the most general $2D$, $N = 2$ action involving scalar and vector multiplets
coupled to supergravity?''  The analogous question for   $4D$, $N = 1$
supergravity
coupled to supermatter was answered a very long time ago and plays a crucial
role in determining the low-energy superstring theory effective action
which in turn provides the starting point in many discussions of
string-inspired
GUT models. As we shall see, $2D$, $N = 2$ theory brings about some new
features in comparison to $4D$, $ N = 1$ theory. To some extent this is
due to the special role of the $2D$ superconformal group. However, another
feature is the fact that $2D$, $N = 2$  models  can realize another symmetry
that presently has no known $ 4D$  field theoretic analogue, namely
 ``mirror symmetry.''  As far as  $2D$  supersymmetric field theory  is
concerned, mirror symmetry has its origins in the simple fact that the
fundamental and simplest representation of $2D$, $ N = 2$ supersymmetry, the
scalar multiplet, comes in two different varieties: there are ``chiral''
multiplets \cite{bible}
and there are ``twisted chiral''  multiplets \cite{GHR}.

It is possible to consider a transformation where a given theory involving
chiral multiplets is mapped into a theory of twisted chiral multiplets
and  vice versa. This is the mirror map. Theories that are
invariant under this map are said to possess mirror symmetry.\footnote{
Although we take the name for this symmetry from its designation in
string theory \cite{mirror}, the \newline ${~~~~}$ discovery of the twisted
chiral multiplet preceeded the discovery of mirror symmetry.}  In addition to
its
applications in string theory, mirror
symmetry is proving to be a  useful tool in
 algebraic geometry and  has
topological implications too.  It is possible that  mirror symmetry plays a
useful role
in the study of some integrable systems  as well.
 Although we will not
study this question, this provides added motivation for us to emphasize
  the mirror-symmetric aspects of the models we consider.

\sect{The Mirror Versions of 2D, N = 2 Supermatter}

   There is a  simple
argument\footnote{This observation is due to T.Hubsch.} to show
the existence of mirror versions
of $2D$, $N = 2$ scalar multiplets,  namely chiral and
twisted chiral multiplets.  One of the
consequences of supersymmetry is  the existence of  equal numbers
of bosons and fermions within a supersymmetric multiplet. In two
dimensions, $N = 2$
scalar multiplets  must  contain two spin-0 bosonic fields and two
spin-1/2 fermionic fields.  The two scalars can be presented as complex
fields, say $\phi$ and ${\Bar \phi}$, which are the
leading components of superfields $\Phi$  and ${\bar \Phi}$. The
spinor components can be defined via application of the superspace spinor
derivatives $D_+$ $D_-$, ${D}_{\pd}$ and ${D}_{\md}$ to either
$\Phi$ or ${\Bar \Phi}$.  Thus, for possible spinor components we
have,
\begin{equation}
\begin{array}{cccc}
D_+ \Phi ~~~,~~~ & D_-  \Phi ~~~,~~~ & {D}_{\pd} \Phi ~~~,~~~ &
{D}_{\md} \Phi ~~~, \\
D_+ {\Bar \Phi} ~~~,~~~ & D_-  {\Bar \Phi} ~~~,~~~ &
{D}_{\pd} {\Bar \Phi}  ~~~,~~~ & {D}_{\md} {\Bar \Phi}  ~~~.
\end{array}
\end{equation}
Since there
are too many spinor states to describe the required two spin-1/2 degrees
of freedom we are forced to set four of the derivatives to zero in order to
obtain
a minimal multiplet.
If we set ${D}_{\pd} \Phi =
{D}_{\md} \Phi = 0$, we find a chiral multiplet. If we set $ {D}_{\pd}
\Phi = D_- \Phi = 0$, we find a twisted chiral multiplet. Finally
if we set ${D}_{\pd}  \Phi = D_+ \Phi = 0$, (breaking $2D$ chiral symmetry)
we find a $2D$, (2,0) righton multiplet.  There are no other choices.\footnote
{From this point of view, the existence of semi-chiral \cite{Buscher}
superfields
which rely on nonlinear constraints is puzzling.} Thus,
simple counting reveals the presence of the twisted chiral multiplet as a
possibility in any $2D$ supersymmetric theory. Of course, nonminimal
representations are possible. For example, the four extra spinors above
may be identified as {\em auxiliary} spinors. There are two ways of doing this,
 leading to the $2D$, $N=2$ complex and twisted complex linear scalar
multiplets. We shall generically denote
chiral superfields by $\Phi$ and twisted chiral superfields  by ${\cal X}$,
satisfying the constraints  $D_{\pd} \Phi = D_{\md}\Phi =0$ and
$D_{\pd}{\cal X}=D_- {\cal X}=0$ respectively, as well as corresponding
conditions for their complex conjugates.

Having defined the two irreducible scalar multiplets through
their respective differential constaints, the component fields
 are defined using the projection method. These
are for the chiral multiplet
\begin{eqnarray}
\F| = \f &,& \Bar{\F} | = \Bar{\f}  \\ \nonumber
D_+ \F | = \psi_+ &,& D_{\pd} \Bar{\F} | = \psi_\pd  \\ \nonumber
D_- \F | = \psi_- &,& D_{\md} \Bar{\F} | = \psi_\md   \\ \nonumber
\sihalf [D_+, D_-] \F| = F &,&\sihalf  [D_{\pd}, D_{\md}] \Bar{\F}| =
\Bar{F} ~~~,
\end{eqnarray}
and for the twisted chiral multiplet
\begin{eqnarray}
{\cal X}| = \chi &,& \Bar{\cal X}| = \Bar {\chi}  \\ \nonumber
D_+ {\cal X} | = \eta_+ &,& D_{\pd} \Bar{\cal X}| = {{\eta}}_\pd  \\
\nonumber
D_{\md} {\cal X}| = \eta_\md &,& D_- \Bar{\cal X} | = {{\eta}}_-
{}  \\ \nonumber
\sihalf[D_+, D_{\md}] {\cal X}| = G &,& \sihalf[D_{\pd}, D_-] \Bar{\cal
X}| = \Bar{G}  ~~~.
\end{eqnarray}

Although we will not study them in detail, we discuss briefly the nonminimal
scalar
multiplets \cite{bible} described by complex linear superfields $\S$ and
twisted
complex linear superfields  $\Xi$, and defined through the second order
differential equations,
\beq
{}~~~~~ D_+ D_- {\Bar \S} ~=~ 0 ~~~~~,~~~~~ D_+ D_{\md} {\Bar \Xi}  ~=~ 0 ~~~~.
\eeq
The components of the non-minimal multiplet $\S$ are given by
\bea
B ~\equiv~  \S \, | &&,~~~  \zeta_\pmd ~\equiv~ D_{\pmd} \S
\, | ~~~, ~~~ \cr
{\rho}_{\pm} ~\equiv~ D_{\pm}  \S \, | &&,~~~ H ~\equiv~ - i
\, D_+ D_- \S \, | ~~~,\cr
u ~\equiv~ - i \, D_\pd D_- \S \, | &&,~~~
v ~\equiv~  - i \, D_\md D_+ \S \, | ~~~, ~~~ \cr
p_{\dpx} ~\equiv~  - i \, D_\pd D_+ \S \, | &&,~~~
p_{\DM} ~\equiv~  - i \, D_\md D_- \S \, | ~~~, ~~~ \cr
\b_{\pmd} ~\equiv~ - i \,  D_+ D_{\pmd}
D_- \S \, | &&,
\eea
with the components of $\Xi$ defined in a similar manner.
We see that many more components exist for the non-minimal
multiplet than for the chiral multiplet.  However, by explicit
evaluation of the free kinetic energy actions  one can show that only $B$,
$\zeta_+$ and $\zeta_-$ are propagating degrees of freedom and this
is exactly the same as for a chiral multiplet.

As in $ 4D$, $N = 1$ superspace, there are several types of superspace
Berezinian
integrals associated with the D-terms and F-terms (and their conjugates). The
D-term superinvariants are of the form
\beq
S_{D} ~=~ \int  d \q^+ d \q^- d{\q}^{\pd} d{\q}^{\md} ~ {\cal L} ~~~,\label{SD}
\eeq
where ${\cal L}$ is any general superfield. The F-term superinvariants
 are obtain from expressions of the form
\beq
S_{F} ~=~ \int  d \q^+ d \q^-  ~ {\cal L}_c ~~~, \label{SF}
\eeq
where ${\cal L}_c$ is a chiral superfield.  However, unlike  $4D$, $ N = 1$
superspace, there is an  additional possibility of forming superinvariants,
namely
\beq
S_{TF} ~=~ \int  d \q^+ d{\q}^{\md} ~ {\cal L}_{tc} ~~~. \label{STF}
\eeq
We can call these ``twisted F-terms'' in analogy with the usual F-terms.
The superfield ${\cal L}_{tc}$ must be a twisted chiral superfield.

These actions can be used to construct in particular, as is well-known,
 nonlinear $\s$-models
involving chiral and twisted chiral superfields. Furthermore, the existence of
linear multiplets gives rise to new possibilities
for the manifest realization of  $2D$ ,  $N = 2$ nonlinear $\s$-models
with torsion.  The simplest linear kinetic term describing the free propagation
of several such
multiplets is given by
\beq
{\cal S} ~=~ \int d^2 x \int d^4 \theta ~ \Big[  ~ - \, {\Bar \S}^i
\, \S^i  ~+~ {\Bar \Xi }^{\rm I}  \,  {\Xi }^{\rm I} ~ \Big] ~~~~.
\eeq
More complicated actions that involve $\a ( \S^i  \S^i  + {\Bar \S}^i {\Bar
\S}^i )$ terms
as well as $\b (  {\Xi }^{\rm I}   {\Xi }^{\rm I}  + {\Bar \Xi }^{\rm I}  {\Bar
\Xi }^{\rm I}  )$  may
also be used. For most values of the constant parameters $\a$ and $ \b$, these
correspond to field redefinitions at the level of component fields.  Clearly,
it is
possible to describe non-linear $\s$-models that also involve ordinary chiral
and twisted chiral supermultiplets by considering a general action of the form
\beq
{\cal S} ~=~ \int d^2 x \int d^4 \theta  ~ {\Hat \O} (\Phi, {\cal X}, \S, \Xi,
\Phib , \Bar{{\cal X}}
, \Bar{\S}, \Bar{\Xi}) ~~~~, \label{sigma}
\eeq
where we regard each type of superfield as providing a coordinate to
describe a subspace of a complex manifold.  A  remarkable point is that there
are now
new manifest  $(2,2)$ supersymmetric ways to  introduce  torsion
on the complex manifold.   We note that the superfields
$D_{\pd} \S$ and $D_{\md} \S$ are  chiral superfields while
$D_{\pd} \Xi$ and $D_- \Xi$ are twisted chiral superfields.  Therefore
it is possible to introduce additional  terms in the action of the form
\bea
{\cal S}' ~=~ &&\int d^2 x \int d^2 \theta ~ [ ~ h_{{\rm J} \, {\rm K} } (\Phi)
 \,
(\, D_{\pd} \S^{\rm J} \,) (\, D_{\md} \S^{\rm K} \,)  ~+~ {\rm{h.\,c.}} ~]
 \non\\
&+&\int d^2 x \int d \theta^+  d \theta^\md ~ [~
 k_{i \, j}  ( {\cal X} ) \,
(\, D_{\pd} \Xi^i \,)  (\,  D_- \Xi^j \,)  ~+~ {\rm {h. \, c.}}    ~]   ~~~~,
\label{add}
\eea
which may be added freely to the non-linear $\s$-model term in \reff{sigma}.

Let us specifically point out what is new about the sum of  \reff{sigma} and
\reff{add}.
For a suitable class of such models, constructed with two multiplets,
 we expect conformal invariance to be
maintained at the quantum level.  As a specific example, let
us retain  one chiral scalar and one non-minimal scalar multiplet. The
target space is then a four-dimensional real manifold and in the usual manner
of string theory possesses a metric, axion and dilaton. In our specific
example, these three target space fields are  determined by ${\Hat \O}$
and $h_{{\rm J} \, {\rm K} }$.  The new feature of this
construction is that previously a single real potential determined all
three target space fields. In these models, the target space fields
are determined by {\em both} ${\Hat \O}$ and $h_{{\rm J} \, {\rm K}}$
and the geometrical constraints of the older models may no longer apply.
Thus we have  additional  $2D$, $N = 2$
 superstrings
beyond the three that were pointed out in 1989 \cite{GLO}.   These new $N = 2$
superstrings
may lead to new classes of exact solutions for $4D$ gravity coupled
to matter.

In a $N=2$ string/conformal field theory context, the feature that
distinguishes chiral and twisted chiral modes is the assignment of their
$U(1)$ charges, and the mirror map is defined as the isomorphism that
interchanges these modes \cite{ALW}. In a lagrangian field theory context
we introduce a
number of chiral multiplets (denoted by $\Phi^i$) and an equal
number of twisted chiral multiplets (denoted by ${\cal X}^i$). An operator
${\cal M}_m$, ``the mirror map operator'', can be introduced through its action
on these fields
\begin{equation}
{\cal M}_m : \Phi^i ~=~ {\cal X}^i  ~~~,~~~ {\cal M}_m : {\cal X}^i ~=~
\Phi^i ~~~. \label{A}
\end{equation}
The mirror map transformation within the
context of supersymmetric $2D ~N = 2$ field theory is somewhat similar
to the mapping of a scalar into a pseudoscalar.

 Let ${\cal L} (\Phi , \, {\cal X}
)$
denote a ``D-term'' superfield Lagrangian that depends on chiral and twisted
chiral superfields. If the equation
\begin{equation}
{\cal M}_m : {\cal L}  ~=~ - \, {\cal L} ~~~, \label{Mm}
\end{equation}
is satisfied, then we say the Lagrangian possesses ``mirror symmetry.''
Two examples of such Lagrangians are provided by,
\begin{equation}
{\cal L}_1 ~=~ \bar{\Phi} {\Phi} ~-~ \bar{\cal X} {\cal X}
{}~~~, ~~~  {\cal L}_2  ~=~  \ln \left[ \frac {\bar{\Phi} {\Phi} }
{ \bar{\cal X} {\cal
X} }   \right]  ~~~~. \label{ln}
\end{equation}

The condition above will lead to an invariant action if
${\cal M}_m$  acts on the spinorial coordinates of
$2D$, $N = 2$ superspace according to the rules
\begin{equation}
{\cal M}_m : \q^+  ~=~ \q^+ ~~~, {\cal M}_m : \q^-  ~=~ {\q}^{\md} ~~~,
{\cal M}_m : {\q}^{\pd}  ~=~ {\q}^{\pd} ~~~,
{\cal M}_m : {\q}^{\md}  ~=~ {\q}^- ~~~.
\end{equation}
This equation acts as the fundamental superspace definition of the
mirror transformation.   Applying this
definition of the mirror operator to the coordinates of the
scalar superfield immediately leads to the results in \reff{A}.
A lagrangian satisfying \reff{Mm} will lead to an action
\beq
S= \int d^2x d^4 \th {\cal L}
\eeq
invariant under mirror
symmetry, since the sign change in the lagrangian is compensated by the
sign change in the measure
$ d \q^+ d\q^- d\q^{\pd} d\q^{\md} \rightarrow
d \q^+  d{\q}^{\md} d{\q}^{\pd} d{\q}^-  = -  \, d \q^+
d \q^- d {\q}^{\pd} d{\q}^{\md}$.

The argument above applies to the invariance of a superfield Lagrangian
that is integrated over the full superspace. It  must be
slightly modified to cover the case of the chiral integrals.  To
achieve invariance of superpotentials we must consider  an action
of the form
\begin{equation}
{\cal S} ~=~ \int d^2 x d \q^+  d{\q}^- ~ U(\Phi^i ) ~+~
\int d^2 x d \q^+  d{\q}^{\md} ~ { U} ({\cal X}^i ) ~+~
{\rm h.}\, {\rm c.} ~~~, \label{pot}
\end{equation}
with the same superpotential function $U$.

\sect{The Mirror Versions of 2D, N = 2 SUSY YM}

Although it
was known some time ago that both chiral and twisted chiral matter
multiplets existed \cite{GHR}, it was not realized until the work of Hull,
Papadopoulos and Spence \cite{HSP} and later Ro\v{c}ek and Verlinde \cite{RV},
that the
mirror transformation can be extended to $2D$, $N = 2$ supersymmetric
vector multiplets.  This can be seen
through the following simple argument:
in all $2D$ theories wherein  a gauge spin-1 field occurs {\em
only} via its field strength, it is possible to perform a duality
transformation to replace the field strength by an auxiliary scalar field.
 This
is the usual Hodge duality transformation in the special case of  $D = 2$.
Consequently, one can regard a $2D$ supersymmetric
vector multiplet as simply an off-shell $2D$ scalar multiplet where one of
the usual auxiliary fields has been replaced by the field strength of a
gauge spin-1 field. This argument applies directly to the two distinct
scalar multiplets (chiral or twisted chiral). Replacing  one of the
auxiliary fields of the chiral matter multiplet by the field strength
of a gauge spin-1 field
yields the $2D$, $N = 2$ ``twisted vector multiplet''.
 Replacing one of the auxiliary fields of a twisted chiral matter multiplet
by the field strength of a gauge spin-1 field yields the
$2D$,  $N = 2$ ``vector multiplet''.

The two irreducible vector multiplets are defined by the
 covariant derivatives $\DelA = D_A + i  \G_A t+ i  {\G'}_A t' $
 which satisfy the constraints
\beq
 \{ \nabla_+ ~,~ \nabla_+ \} ~=~ 0 ~~~,~~~ \{ \nabla_- ~,~ \nabla_- \} ~=~ 0
{}   $$
$$ \{ \nabla_+ ~,~ { \nabla}_- \} ~=~ - i  {\Bar {\cal P}} \, t'
{}~~~, ~~~ \{ \nabla_+ ~,~ {\nabla_{\md}} \} ~=~ i  {\Bar {\cal W}} \, t
{} $$
$$ \{ \nabla_+ ~,~ {\nabla}_{\pd} \} ~=~ i  \nabla_{\dpx} ~~~,~~~
\{ \nabla_- ~,~ {\nabla}_{\md} \} ~=~ i  \nabla_{\DM}  $$
$$ [ \nabla_+ ~,~ \nabla_{\dpx} ] ~=~ 0 ~~~,~~~ [ \nabla_- ~,~ \nabla_{\DM} ]
 ~=~ 0    $$
$$ [ \nabla_+ ~,~ \nabla_{\DM} ] ~=~ -\,   ({\nabla}_- {\Bar {\cal W}} ) \, t
{}~+~   ({\nabla}_{\md} {\Bar {\cal P}} ) \, t'  $$
$$ [ \nabla_- ~,~ \nabla_{\dpx} ] ~=~    ({\nabla}_+ { {\cal W}} ) \, t
{}~+~   ({\nabla}_{\pd} {\Bar {\cal P}} ) \, t'
{} $$
$$ [ \nabla_{\dpx} ~,~ \nabla_{\DM} ] ~=~ i  {\cal F} \, t ~+~ i  {\cal F}' \,
t'
 ~~~,  {~~~~}
\eeq
and the relations implied by the Bianchi identities.
Here $t$ and $t'$ are Lie
algebra generators of two distinct $U(1)$ groups (the generalization to
other Lie algebras is simple).
It is clear that under the mirror map we have for vector multiplets
\begin{equation}
{\cal M}_m : {\cal P} ~=~ {\cal W}  ~~~,~~~ {\cal M}_m : {\cal W} ~=~
{\cal P} ~~~.
\end{equation}

The action of  the Lie algebra generators  on {\em covariantly}
chiral matter fields $\F$ or twisted chiral matter
fields ${\cal X}$  is restricted by an integrability condition, (e.g. $ 0=
\{\Delpd ,\Delmd \}
\Phi = -i \bar{{\cal P}} t' \Phi)$ so that
\bea
&&[~ t \, , \,  \Phi ~] ~=~ iq \Phi ~~~,~~~ [~ t' \, , \, {\cal X}~] ~=~
iq' {\cal X}  \nonumber \\
&&[~ t' \, , \, \Phi ~] ~=~ 0~~~~~~,~~~ [~ t \, , \, {\cal X}~] ~=~ 0
{}~~~,
\eea
where $q$, $q'$ are the $U(1)$ charges of the multiplets.
The components of the twisted vector multiplet are given by
\bea
{\cal P}| ~=~ P &,&  \Bar{\cal P}|  ~=~ \Bar{P}   \\ \nonumber
\Delp {\cal P}| ~=~ \rho_+ &,& \Delpd \bar{\cal P} |  ~=~ \rho_\pd   \\
\nonumber
\Delm {\cal P}| ~=~ \rho_- &,& \Delmd \bar{\cal P} | ~=~ \rho_\md  \\
\nonumber
\sihalf[\Delp, \Delm] {\cal P}| ~=~ { H} &,& \sihalf[\Delpd, \Delmd]
\bar{\cal P}| ~=~ {\Bar { H}}  ~~~,
\eea
while those of the vector multiplet are given by
\bea
{\cal W}| = W &,& \Bar{\cal W}| = \Bar{W}  \\ \nonumber
\Delp {\cal W}| = \l_+ &,& \Del_{\pd} \Bar{\cal W}| = {{\l}}_\pd
{} \\ \nonumber
\Delmd {\cal W}| = \l_\md &,& \Delm \Bar{\cal W} | = {{\l}}_-
{}  \\ \nonumber
\sihalf[\Delp, \Del_{\md}] {\cal W}| = { J}   &,& \sihalf[\Del_{\pd}, \Delm]
\Bar{\cal W}| = {\Bar { J}} ~~~.
\eea
The actual gauge fields occur through their respective field strengths inside
 the quantities $H$  and $J$ above, as  seen by writing $H$ and $J$ in
terms of their real and imaginary parts
\beq
H=\half [{\rm d}' +i {\rm F}(A')] ~~~,~~~J=\half[{\rm d}+i{\rm F}(A)]  ~~
\eeq
 where ${\rm F}(A) = \pa_{\pp} A_{\mm} -  \pa_{\mm} A_{\pp}$
 and  ${\rm F}(A') = \pa_{\pp}
A_{\mm}' - \pa_{\mm} A_{\pp}'$.

Comparing the component projection equations
for the scalar multiplets with those for the vector multiplets shows
quite clearly that one of the auxiliary fields in the former is replaced
by a spin-1 field strength in the latter.

In closing, it should be  mentioned that by the duality argument presented
 above there are also
more vector multiplets and twisted vector multiplets since  both
the chiral and twisted chiral matter superfields have {\em two}
auxiliary fields.  We can choose to replace {\em both} auxiliary
fields by the
field strengths of gauge spin-1 vectors.  Thus, it is possible to have
irreducible $2D$, $N = 2$ multiplets that have two gauge fields in the
same multiplet.  Obviously, there exist also distinct vector multiplets
obtained
by applying a duality transformation to auxiliary spin-zero fields of complex
linear
or twisted complex linear superfields.  (This
same phenomenon occurs in the description of $2D$, $N = 2$ supergravity.
 One can replace some or all of the auxiliary
fields that occur in the off-shell theory by field strengths of gauge
spin-1 vectors,  leading to many distinct off-shell versions of $2D$,
$N = 2$ supergravity.)

\sect{The Mirror Versions of 2D, N = 2 Supergravity}

In the previous sections we saw that matter scalar and vector
multiplets  realizing $2D$, $N = 2$ supersymmetry come in mirror realizations.
In the same manner, irreducible $N=2$ supergravity comes in mirror versions,
the so-called $U_V(1)$ and $U_A(1)$ versions \cite{HP,GLO,MGMW}. Mirror
symmetry can be
realized in $N=2$ supergravity by considering the {\em reducible}
$U_V(1) \otimes U_A(1)$ theory \cite{HP,MGMW}.\footnote{This
again suggests that it may actually
be possible to find remnants of mirror symmetry in \newline ${~~~~}$
supersymmetric integrable systems.}

The mirror symmetric form of $2D$,   $U_V (1) \otimes U_A (1)$ $N = 2 $
supergravity
is described in  ref. \cite{HP} and further discussed in
ref. \cite{MGMW}.
Its tangent space symmetries are Lorentz, $U_V(1)$ and $U_A(1)$. Their
respective generators  (with a slight change of notation from the above
references) are
denoted by ${\cal M}$, ${\cal Y}$ and
${\cal Y}'$. Their  action on spinors is
\bea
[~{\cal M} \, , \, \psi_{\pm}~] ~=~ \pm \frac{1}{2} \psi_{\pm} ~~~~&,&~~~~
[~{\cal M} \, , \, \psi_{\pmd}~] ~=~ \pm \frac{1}{2} \psi_{\pmd}
\nonumber\\
{[}~{\cal Y} \, , \,  \psi_{\pm}~] ~=~  - \frac{i}{2} \psi_{\pm} ~~~~&,&~~~~
[~ {\cal Y} \, , \, \psi_{\pmd}~] ~=~ +  \frac{i}{2} \psi_{\pmd}
\nonumber\\
{[}~ {\cal Y}' \, , \, \psi_{\pm}~] ~=~ \mp  \frac{i}{2} \psi_{\pm}~~~~&,&~~~~
[~ {\cal Y}' \, , \, \psi_{\pmd}~ ]  ~=~ \pm  \frac{i}{2} \psi_{\pmd} ~~~  .
\ena
We also define the combinations
\bea
M &=& \frac{1}{2}({\cal M} +i {\cal Y} ')~~~~,~~~
 \bar{M}= \frac{1}{2}({\cal M} -i {\cal Y} ')\nonumber \\
N &=& \frac{1}{2}({\cal M} +i {\cal Y} )~~~~,~~~
 \bar{N}= \frac{1}{2}({\cal M} -i {\cal Y} )~~.
\eea
The covariant derivatives are defined by
\bea
\Hat{\Del}_A  &=& E_A + \F_A {\cal M} + \S_A'  {\cal Y}'+ \S_A {\cal Y} ~~.
\eea
 The constraints which define the $2D$,  $N = 2$  $U_V (1) \otimes U_A (1)$
supergravity are given by
\beq
 \{ {\Hat \nabla}_+ ~,~ {\Hat \nabla}_+ \} ~=~ 0 ~~~,~~~ \{ {\Hat \nabla}_-
{}~,~ {\Hat \nabla}_- \} ~=~ 0
{}~~~, ~~~ \{ {\Hat \nabla}_+ ~,~ {\Hat \nabla}_{\md} \} ~=~ - \,
{\Bar F}  \, {\Bar N} ~~~,  $$
$$ \{ {\Hat \nabla}_+ ~,~ {\Hat  \nabla}_- \} ~=~ - \,  {\Bar R} \,
{\Bar M}
{}~~~, $$
$$ \{ {\Hat \nabla}_+ ~,~ {\Hat \nabla}_{\pd} \} ~=~ i  {\Hat \nabla}_{\dpx}
 ~~~,~~~ \{ {\Hat \nabla}_- ~,~ {\Hat \nabla}_{\md} \} ~=~ i
{\Hat \nabla}_{\DM} ~~~ .
\eeq
In writing this, we have used a ``hatted'' symbol for the $U_V(1) \otimes
U_A(1)$ covariant derivatives to distinguish them from the irreducible
derivatives below.

{}From these it follows that
\bea
{[}~{ \Hat{\Del}}_{+} \, , \, {\Hat {\Del}}_{\pp} ~] &=& 0 ~~~, ~~~[~
{ \Hat{\Del}}_{-} \, , \, { \Hat{\Del}}_{\mm} ~] = 0
{}~~~, \\ \non
{[}~{ \Hat{\Del}}_{\pd} \, , \, { \Hat{\Del}}_{\pp} ~] &=& 0 ~~~, ~~~[~
{ \Hat{\Del}}_{\md} \, , \, { \Hat{\Del}}_{\mm} ~] =
0 ~~~, \\ \non
{[}~{ \Hat{\Del}}_{+} \, , \, { \Hat{\Del}}_{\mm} ~] &=& - \sihalf \Rb
{ \Hat{\Del}}_{\md}  -i({ \Hat{\Del}}_{\md} \Rb) \Mb
 - \sihalf \bar{F} \Hat{\Del}_-
-i ({ \Hat{\Del}}_-  \bar{F} ) \bar{N}
{}~~~, \\ \non
{[}~{ \Hat{\Del}}_{\pd} \, , \, { \Hat{\Del}}_{\mm} ~] &=&  \sihalf R
{\Hat {\Del}}_{-}  +i( {\Hat {\Del}}_- R) M
+ \sihalf F
{\Hat {\Del}}_{\md}  +i( {\Hat {\Del}}_{\md} F) N ~~~, \\ \non
{[}~{ \Hat{\Del}}_{-} \, , \, { \Hat{\Del}}_{\pp} ~] &=&  \sihalf \Rb
{ \Hat{\Del}}_{\pd}  -i({ \Hat{\Del}}_{\pd} \Rb)\Mb
 +\sihalf F
{ \Hat{\Del}}_{+}  -i({ \Hat{\Del}}_{+} F)N
{}~~~, \\ \non
{[}~{ \Hat{\Del}}_{\md} \, ,\,  { \Hat{\Del}}_{\pp} ~] &=& - \sihalf R
{ \Hat{\Del}}_{+}  +i({ \Hat{\Del}}_{+} R) M
-\sihalf \bar{F}
{ \Hat{\Del}}_{\pd}  +i({ \Hat{\Del}}_{\pd} \bar{F}) \bar{N}
 ~~~,
\eea
and also
\bea
{[}~ { \Hat{\Del}}_{\pp} \, , \, { \Hat{\Del}}_{\mm} ~] &=&  \half (
{ \Hat{\Del}}_+ R) {\Hat {\Del}}_- +\half ({\Hat {\Del}}_-  R) { \Hat{
\Del}}_+  - \half ({ \Hat{\Del}}_{\pd} \Rb) {\Hat {\Del}}_{\md} - \half (
{ \Hat{\Del}}_{\md} \Rb) {\Hat {\Del}}_{\pd} \non \\
&& -  \half R \Rb \Mb - \half R \Rb M + ({{ \Hat{\Del}}}^2 R) M -(
{{ \Hat{\Delb}}}^2 \Rb) \Mb  \non\\
&& + \half (
{ \Hat{\Del}}_+ F) {\Hat {\Del}}_{\md} +\half ({\Hat {\Del}}_{\md} F)
 { \Hat{
\Del}}_+  - \half ({ \Hat{\Del}}_{\pd} \bar{F}) {\Hat {\Del}}_{-}
 - \half (
{ \Hat{\Del}}_{-} \bar{F}) {\Hat {\Del}}_{\pd} \non \\
&& -  \half F \bar{F} \bar{N} - \half F \bar{F} N +
({ \Hat{\Del}}_+ { \Hat{\Del}}_{\md} F) N -(
{ \Hat{\Del}}_{\pd} { \Hat{\Del}}_-  \bar{F}) \bar{N}
 ~~.
\eea

The two distinct irreducible forms of $2D$, $N = 2$ supergravity are obtained
by restricting the gauge group so that either
$F=0$ for the $U_A(1)$ version,
or $R=0$ for the $U_V(1)$ version,
 so that the
corresponding connections  $\Sigma_{A}$ or ${\S}_{A}'= \Omega_{A}
 - \Gamma_{A}$ are pure gauge and can be removed.
Thus, the $U_V (1)$ theory is  described by
\beq
 \{ \nabla_+ ~,~ \nabla_+ \} ~=~ 0 ~~~,~~~ \{ \nabla_- ~,
{}~ \nabla_- \} ~=~ 0
{}~~~, ~~~ \{ \nabla_+ ~,~ {\nabla_{\md}} \} ~=~-\Bar{F} \, \Bar{N}
{}~~~,  $$
$$ \{ \nabla_+ ~,~ { \nabla}_- \} ~=~ 0
{}~~~, $$
$$ \{ \nabla_+ ~,~ {\nabla}_{\pd} \} ~=~ i  \nabla_{ \pp} ~~~,~~~
\{ \nabla_- ~,~ {\nabla}_{\md} \} ~=~ i  \nabla_{\DM} ~~~ , \label{UV}
\eeq
while the $U_A (1)$ theory is described by
\beq
 \{ \nabla_+ ~,~ \nabla_+ \} ~=~ 0 ~~~,~~~ \{ \nabla_- ~,~ \nabla_- \} ~=~ 0
{}~~~, ~~~ \{ \nabla_+ ~,~ {\nabla_{\md}} \} ~=~ 0 ~~~,  $$
$$ \{ \nabla_+ ~,~ { \nabla}_- \} ~=~- \Rb \Mb
{}~~~, $$
$$ \{ \nabla_+ ~,~ {\nabla}_{\pd} \} ~=~ i  \nabla_{ \pp} ~~~,~~~
\{ \nabla_- ~,~ {\nabla}_{\md} \} ~=~ i  \nabla_{ \mm} ~~~ , \label{UA}
\eeq
and their consequences.
However,
 setting either ${\Bar F}$ or ${\Bar R}$ to zero {\it {breaks}} mirror
symmetry.
Thus, in order to construct a local $2D$, $N = 2$ , mirror symmetric,
 supersymmetric theory, we
must work with the reducible representation in (4.4).

Let us note that there are precedents for working with a reducible
supergravity representation.  The most important of these is the $4D$ ,
$N = 1 $ limit of heterotic string theory.  It is known that in this limit
the supergravity multiplet appears along with a specific
matter multiplet (the linear or tensor multiplet) that contains the
axion, dilaton and dilatino.  Thus, the low energy limit of the
heterotic string is reducible with respect to supersymmetry.
However, it forms an irreducible representation of some
larger symmetry present in string theories.
The reducible representation described
by (4.4) is the analogue of the supergravity plus tensor
multiplets seen in the low energy limit of superstring theory.

In \cite{measures} we have shown how  to construct the covariant
derivatives of the   $U_V (1) \otimes U_A (1)$  theory in terms of those of the
 $U_A(1)$ theory and the prepotentials  that solve the supergravity
constraints.
An equivalent procedure consists of
starting  with the supergravity covariant derivatives that solve the
constraints of the $U_A(1)$ covariant derivative in (4.8),
 and ``entangle''  them with
a matter twisted vector multiplet. We define  new covariant derivatives
through
\bea
{\Hat \Del}_{+} &=& \exp{[ \frac 12 (k + l) V]} \, [~ {\Del}_+ ~+~ (\,
{\Del}_+ V \,) (\, - (k + m) {\cal M} \,+\, i k {\cal Y}
\,+\, i k {\cal Y}' \, )
{}~ ]   \non\\
{\Hat \Del}_{-} &=& \exp{[ \frac 12 (k - l) V]} \, [~ {\Del}_- ~+~ (\,
{\Del}_- V \,) (\, (k - m) {\cal M} \,+\, i k {\cal Y}
\,+\, i k {\cal Y}' \, ) ~ ] \label{ent}
\eea
where $V$ is an arbitrary real scalar superfield and $k, \, l \,$
and $m$ are real parameters.
 We note that since we are dealing with an $N=2$ theory, the scale
transformation represented by the exponential factor on the right-hand-side of
these equations does not lead to a rescaling of the vielbein determinant (as
would be
the case for other values of $N$), and hence
\beq
\Hat{E}^{-1} = E^{-1}~~.
\eeq

The commutator algebra of the covariant derivatives thus defined is isomorphic
to the commutator algebra of the $U_V (1) \otimes U_A (1)$
supergravity covariant derivative. In particular, the field strengths are given
by
\beq
{\Bar F}_{U_V \otimes U_A } = 4k \, \exp[ - k V ]  \Big[ \, {\Del}_+
{\Del}_{\md}
 V \, \Big]
{}~~~,~~~ {\Bar R}_{U_V \otimes U_A } = \exp[ - k V ] \Big[ {\Bar R}_{U_A } ~-~
4 k (
 {\Del}_+ {\Del}_- V \, ) ~  \Big] ~~~ .
\eeq
An equivalent result is obtained if one starts from the irreducible $U_V (1)$
supergravity covariant derivative and entangles properly with a (untwisted)
vector multiplet.

\sect{ Local Integration  in 2D, N = 2 Superspace}

In this section we summarize the  information concerning the
local supermeasures for generalizing the global invariants in
\reff{SD}-\reff{STF} in the presence of supergravity. We are primarily
interested in the projection formulae for obtaining component
actions from the corresponding superspace actions.

In  local supersymmetry,  the superinvariant given by a
full superspace integral  involves the
vielbein superdeterminant $E$:
\beq
S_{D} ~=~ \int  d \q^+ d \q^- d{\q}^{\pd} d{\q}^{\md} ~
E^{-1}\,  {\cal L} ~~~,
\eeq
where ${\cal L}$ is a general superfield.  This expression is valid for either
of the
minimal theories, or the non-minimal one, {\em with the same} $E$.

Corresponding to the global F-term invariants one  finds
local expressions of the form
\beq
S_{F} ~=~ \int  d \q^+ d \q^-  ~ {\cal E}^{-1}\, {\cal L}_C
\eeq
and
\beq
S_{TF} ~=~ \int  d \q^+ d \q^{\md}  ~ {\Tilde {\cal E}}^{-1} \, {\cal
L}_{TC} ~~~,
\eeq
where ${\cal L}_C  $ and ${\cal L}_{TC}$ are {\em covariantly} chiral and
twisted chiral superfields and $ {\cal E}^{-1}$ and
$ \Tilde {\cal E}^{-1}$ are suitable measures. For $U_A(1)$ supergravity, which
contains the
chiral compensator $\s$ in addition to the (vector) prepotential $H^a$,
the chiral measure is given explicitly by \cite{measures}
\beq
{\cal E}^{-1}=  e^{-2\s} (1.e^{\BAH})
\eeq
and a covariantly chiral ${\cal L}_C$ is given in terms of an
ordinary chiral ${\cal L}_c$ by
\beq
{\cal L}_C = e^H {\cal L}_c e^{-H}
\eeq
or by ${\cal L}_C = \Delb ^2 {\cal L}$ in terms of a general superfield.
In the $U_V(1)$ theory similar results hold for the mirror-mapped
quantities, i.e. twisted chiral measure, etc. (We are using the notation
$H= i H^m \pa_m$ where in $\BAH$ the derivative acts on everything to its
left.)

In \cite{measures}, we have established the existence of the measures
${\cal E}^{-1}$, $\tilde{\cal E}^{-1}$ in the $U_V(1) \otimes U_A(1)$ theory,
and given an explicit formula for the former.  However for our purpose,
the existence of the full
superspace measure $E^{-1}$ and of the chiral measure ${\cal E}^{-1}$,
along with the projection formulae, will suffice.

Component actions are obtained from superspace actions
by projection formulae.
We consider here the $U_A(1)$ version of supergravity described by the vector
prepotential $H^a$ and the chiral scalar prepotential $\s$. (In subsection 7.5
we will discuss generalizations for the   $U_V (1) \otimes U_A (1)$ theory.)
In order to go from a full superspace integral to a chiral one we use
the formula (for a general $\cal L$)
\beq
\int d^2 x d^4 \th E^{-1} {\cal L} = \int d^2 x d^2 \th  {\cal E}^{-1}
\bar{\Del}^2 {\cal L}|_{\bar{\th}=0} ~~.   \label{chint}
\eeq
This superspace integral can be reduced to a component
action by means of the chiral density projection formula \cite{measures}
\bea
&& \int d^2 x d^4 \th E^{-1} {\cal L}  \non \\
&&=  \int d^2 x e^{-1} \left[ \Del^2 +i \psi_=^{\md}\Del_+ -i
\psi_{\pp}^{\pd}\Del_-
+(- \half \bar{B} -\psi_{\pp}^{\md}\psi_=^{\pd}
+\psi_=^{\md}\psi_{\pp}^{\pd})\right]\bar{\Del}^2
{\cal L}|
\eea  \label{chiral density}
where $\psi_a^{\a}$ is the gravitino field and $ \Bar{R}| =\Bar{B} $.

Since, by construction, the expression on the right-hand-side of
this equation is invariant,
we can replace $\Delb^2 {\cal L}$ by any covariantly chiral lagrangian
${\cal L}_C$ and obtain the component projection formula for
any F-type local invariant:
\bea
&&\int d^2 x d^2 \th  {\cal E}^{-1} {\cal L}_C  \nonumber \\
&&= \int d^2 x e^{-1} \left[ \Del^2 +i \psi_=^{\md}\Del_+ -i
\psi_{\pp}^{\pd}\Del_-
+(- \half \bar{B} -\psi_{\pp}^{\md}\psi_=^{\pd}
+\psi_=^{\md}\psi_{\pp}^{\pd})\right]
{\cal L}_C| ~.\label{chproj}
\eea

By straightforward
 but non-trivial algebra \cite{measures}, the general component
 expression can be rewritten in terms of twisted chiral
projectors given by
\bea
&& \int d^2 x d^4 \th E^{-1} {\cal L}  \non \\
&&=\int d^2 x  e^{-1}[ \Del_{\md}\Del_+ +i \psi^{\pd}_{\pp} \Del_{\md}
-i\psi^{-}_{\mm} \Del_{+} +(\psi^{\pd}_{\pp}\psi^-_{\mm}
+\psi^-_{\pp}\psi^{\pd}_{\mm})]
\Del_{\pd}\Del_{-} {\cal L}| ~.\label{twist density}
\eea
Although in the $U_A(1)$ theory we do not know an explicit expression for
 $ {\Tilde {\cal E}}^{-1}$, it is reasonable to assume that
replacing $\Del_{\pd}\Del_{-} {\cal L}$ by any covariantly twisted
chiral lagrangian ${\cal L}_{TC}$ would allow us to project to
components a corresponding twisted  F-term according to
\bea
&& \int d^2 x d \q^+ d \q^{\md}  ~ {\Tilde {\cal E}}^{-1} \, {\cal
L}_{TC}  \nonumber\\
&&=-\int d^2 x  e^{-1}[ \Del_{\md}\Del_+ +i \psi^{\pd}_{\pp} \Del_{\md}
-i\psi^{-}_{\mm} \Del_{+} +(\psi^{\pd}_{\pp}\psi^-_{\mm}
+\psi^-_{\pp}\psi^{\pd}_{\mm})]
 {\cal L}_{TC}| ~.
\eea
We are now in a position to construct component lagrangians for a
variety of models, but let us conclude this section with the following
remark:

In $D=4$, it is known that the transition from global to local theory
for  a general K\"{a}hler action $\int d^4 \th K( \Phi , \bar{\Phi})$,
invariant under K\"{a}hler transformations $K \rightarrow K+ \Lambda_c
+\bar{\Lambda}_a$ ($\bar{D} \Lambda_c = D \bar{ \Lambda}_a =0$), requires
replacing the global action by the local action
\beq
{\cal S} =\int d^4 x d^4 \th E^{-1} e^K
\eeq
simply because under the K\"{a}hler transformation
$K \rightarrow K+ \Lambda_C
+\bar{\Lambda}_A$ ($\bar{\Del} \Lambda_C = \Del \bar{ \Lambda}_A =0$),
the variation of ${\cal S} = \int d^4 x d^4 \th E^{-1} K $
\bea
\d {\cal S} &=& \int d^4 \th E^{-1} ( \Lambda_C+\bar{\Lambda}_A) \nonumber\\
&=& \int d^2 \th \s^3 ( \Delb ^2 +R) \bar{\Lambda}_A+
\int d^2 \bar{\th} \bar{\s}^3 ( \Del ^2 +\bar{R}) \Lambda_C ~~,
\eea
does not vanish (but in the exponential it can be cancelled by
scale transformations of the compensator $\s$). However, in $D=2$
 the projection formula \reff{chint} to (anti-) chiral
integrals does not
involve the superfield strength $R$ and there is no need to exponentiate
the K\"{a}hler potential.

\sect{General Coupling of Scalar and Vector Multiplets}

In this section we consider the coupling of matter scalar and vector
multiplets.
To begin with we take a general function of  covariantly chiral and twisted
chiral (with respect to the $U(1)$ groups) superfields
\beq
{\cal S} =\int d^2x d^4 \theta K(\Phi , \Bar{\Phi}, {\cal X}, {\Bar{\cal X}})
{}~~~, \label{KFX}
\eeq
in the presence of vector and twisted vector multiplets and evaluating
at $\th = 0$.

We take this down to components by  replacing $d^4 \th$ by
$ \Del_+ \Del_- \Del_{\pd} \Del_{\md}$~, in this order. Here the covariant
derivatives satisfy the constraints in (3.1). After distributing
the four   derivatives,  but for ease of reading {\em suppressing
everywhere  the derivatives of $K$
with respect to the superfields}
 (  factors of $K$ and its
derivatives with respect to the fields can be easily
reinstated by counting the number, and types of the fields which appear - i.e.
a term such as $\Delmd \Phib \Delpd \bar{\calX} \Delm \Phi
\Delp \calX$  should be
 multiplied by the factor $K_{\Phib \bar{\calX} \Phi \calX}$) , we obtain :
\bea
\lefteqn {{\cal S} = \int d^2x  \{ (\Delp \Delm \Delpd \Delmd \Phib +
 \Delp \Delm \Delpd \Delmd \calX)} \non \\
&& - (\Delm \Delpd \Delmd \Phib +\Delm \Delpd \Delmd \calX)
       (\Delp \Phi + \Delp \calX)  \non \\
&& + (\Delp \Delpd \Delmd \Phib + \Delp \Delpd \Delmd \calX)(\Delm \Phi
       + \Delm \bar{\calX} ) \non \\
&&+ (\Delpd \Delmd \Phib + \Delpd \Delmd \calX)
        (\Delp \Delm \Phi + \Delp \Delm \bar{\calX}) \non \\
&& - (\Delpd \Delmd \Phib + \Delpd \Delmd \calX)(\Delm \Phi + \Delm
\bar{\calX})
      (\Delp \Phi + \Delp \calX) \non \\
&& - (\Delp \Delm \Delmd \Phib + \Delp \Delm \Delmd \calX)(\Delpd \Phib
     + \Delpd \bar{\calX}) \non \\
&& - (\Delm \Delmd \Phib + \Delm \Delmd \calX)(\Delp \Delpd \Phib +
      \Delp \Delpd \bar{\calX}) \non \\
&& + (\Delm \Delmd \Phib + \Delm \Delmd \calX)(\Delpd \Phib + \Delpd
\bar{\calX})
      (\Delp \Phi + \Delp \calX) \non \\
&& + (\Delp \Delmd \Phib + \Delp \Delmd \calX)(\Delm \Delpd \Phib
       + \Delm \Delpd \bar{\calX}) \non \\
&& - (\Delmd \Phib + \Delmd \calX)(\Delp \Delm \Delpd \Phib
       + \Delp \Delm \Delpd \bar{\calX}) \non \\
&& - (\Delmd \Phib + \Delmd \calX)(\Delm \Delpd \Phib + \Delm \Delpd
\bar{\calX})
        (\Delp \Phi + \Delp \calX) \non \\
&& - (\Delp \Delmd \Phib + \Delp \Delmd \calX)(\Delpd \Phib + \Delpd
\bar{\calX})
      (\Delm \Phi + \Delm \bar{\calX}) \non \\
&& + (\Delmd \Phib + \Delmd \calX)(\Delp \Delpd \Phib + \Delp \Delpd
\bar{\calX})
       (\Delm \Phi + \Delm \bar{\calX}) \non \\
&& - (\Delmd \Phib + \Delmd \calX)(\Delpd \Phib + \Delpd \bar{\calX})
       (\Delp \Delm \Phi + \Delp \Delm \bar{\calX})  \\
&& + (\Delmd \Phib + \Delmd \calX)(\Delpd \Phib + \Delpd \bar{\calX})
         (\Delm \Phi + \Delm \bar{\calX}) (\Delp \Phi + \Delp \calX)  \}
|\non~~.
\eea

In the reduction to components, we use the  (twisted) chirality conditions
and the (anti-)commutators of  (3.1) to write
\bea
\Delp \Delm \Delpd \Delmd \Phib | &=& \{\shalf (\Delpp \Delmm + \Delmm \Delpp)
  - \squart q [\Delp, \Delmd] {\cal W}
 - \squart q [\Delpd, \Delm] \bar{\cal W} \non \\
&&+q(\Delm \bar{\cal W})\Delpd - q (\Delp {\cal W} )\Delmd
-q^2 {\cal W} \bar{ \cal W} \}\Phib| \non \\
&=& \Box \phib + \sihalf q J \phib +\sihalf q\bar{J} \phib
     + q \l_- \psi_\pd  - q \l_+ \psi_\md - q^2 W\bar{W} \phib  \non\\
\Delp \Delm \Delpd \Delmd \calX | &=& - \shalf q'([\Delp , \Delm]\calP)\calX|
  + q'(\Delm \calP) \Delp \calX | \non \\
&=&  iq'H \chi + q'\rho_- \eta_+ \non \\
\Delm \Delpd \Delmd \Phib| &=& -i \Delmm \Delpd \Phib| -
 q (\Delmd {\cal W})
\Phib|
  -q {\cal W}(\Delmd \Phib)| \non \\
&=& -i \calDmm \psi_\pd - q \l_\md \phib  - q W \psi_\md  \non\\
\Delm \Delpd \Delmd \calX | &=& - q' (\Delm \calP)\calX | \non\\
&=& - q'\rho_- \chi  \non\\
\Delp \Delpd \Delmd \Phib | &=& i \Delpp \Delmd \Phib| -q(\Delpd \bar{\cal W})
\Phib|
  -q \bar{\cal W}(\Delpd \Phib)| \non \\
&=& i \calDpp \psi_\md - q\l_\pd \phib - q\bar{W} \psi_\pd  \non\\
\Delp \Delpd \Delmd \calX | &=& -q' (\Delp \calP)\calX | -q'\calP\Delp \calX |
 \non \\
&=&  -q'\rho_+ \chi -q'P\eta_+  \non\\
\Delp \Delm \Delmd \Phib | &=& -q (\Delm \bar{\cal W})\Phib | \non \\
&=& - q \l_- \phib  \non\\
\Delp \Delm \Delmd \calX | &=& -q'(\Delmd \calPb)\calX |+ i \Delmm \Delp \calX
|
  \non \\
&=& - q'\rho_\md \chi + i \calDmm \eta_+  \non\\
\Delp \Delm \Delpd \Phib | &=& - q (\Delp {\cal W}) \Phib | \non \\
&=&  -q \l_+ \phib  \non\\
\Delp \Delm \Delpd \bar{\calX} | &=& -i \Delpp \Delm \bar{\calX}|
     -i(\Delpd \calPb)t' \bar{\calX}| -i \calPb t'\Delpd
\bar{\calX}| \non \\
&=& - i\calDpp \eta_- - q'\rho_\pd \bar{\chi} - q'\Pb \eta_\pd \non \\
\Delpd \Delmd \Phib | &=& -i\Fb \non \\
\Delp \Delm \Phi | &=& -iF  \non\\
\Delp \Delmd \Phib | &=& q \bar{W} \phib \non \\
\Delm \Delpd \Phib | &=& - q W \phib  \non\\
\Delm \Delp \Phi | &=& i F  \non\\
\Delpd \Delm \Phi | &=& q W \phi  \non\\
\Delmd \Delp \Phi | &=& - q \bar{W} \phi  \non\\
\Delmd \Delpd \Phib | &=& i \Fb  \non\\
\Delp \Delm \bar{\calX} | &=& - q' \Pb \bar{\chi}  \non\\
\Delpd \Delmd \calX | &=&  -q' P \chi  \non\\
\Delp \Delmd \calX | &=& -iG  \non\\
\Delm \Delpd \bar{\calX} | &=& i \bar{G}  \non\\
\Delm \Delp \calX | &=& q' \Pb \chi \non\\
\Delpd \Delm \bar{\calX} | &=&-i \bar{G}  \non\\
\Delmd \Delp \calX | &=& i G   \non\\
\Delmd \Delpd \bar{\calX} | &=&  q'P \bar{\chi} ~~.
\eea

We obtain
 the ``raw" component lagrangian \bea
&& (\Box \phib + \sihalf q J \phib + \sihalf q \bar{J} \phib
+ q \l_- \psi_\pd
     - q \l_+ \psi_\md - q^2 W\bar{W} \phib +iq'H \chi - q'\rho_- \eta_+ ) \non
\\
&& + (i \calDmm \psi_\pd + q \l_\md \phib  +q W \psi_\md +q'\rho_- \chi)
     (\psi_+ + \eta_+) \non \\
&& + (i \calDpp \psi_\md - q\l_\pd \phib - q\bar{W} \psi_\pd - q'\rho_+ \chi
      -q'P\eta_+) (\psi_- + \eta_-) \non \\
&& - (\Fb -i q'P \chi)(F -i q'\Pb \bar{\chi} ) \non \\
&& +i (\Fb -i q'P \chi)(\psi_- + \eta_-)(\psi_+ + \eta_+) \non \\
&& + (q \l_- \phib + q'\rho_\md \chi - i \calDmm \eta_+)(\psi_\pd + \eta_\pd)
    \non \\
&& + (\calDmm \phib + \calDmm \chi)(\calDpp \phib + \calDpp \bar{\chi} ) \non
\\
&& + i(\calDmm \phib + \calDmm \chi)(\psi_\pd + \eta_\pd)(\psi_+ + \eta_+)
   \non \\
&& - (q \bar{W} \phib -i G)(q W \phib -i \bar{G} ) \non \\
&& + (\psi_\md + \eta_\md)(q \l_+ \phib + i\calDpp \eta_- + q'\rho_\pd
\bar{\chi}
       + q'\Pb \eta_\pd) \non \\
&& + (\psi_\md + \eta_\md)(q W \phib -i \bar{G})(\psi_+ + \eta_+) \non \\
&& - (q \bar{W} \phib -i G)(\psi_\pd + \eta_\pd)(\psi_- + \eta_-) \non \\
&& + i (\psi_\md + \eta_\md)(\calDpp \phib + \calDpp \bar{\chi} )(\psi_- +
\eta_-)
   \non \\
&& +i(\psi_\md + \eta_\md) (\psi_\pd + \eta_\pd)(F -i q'\Pb \bar{\chi}) \non \\
&& +(\psi_\md + \eta_\md) (\psi_\pd + \eta_\pd)(\psi_- + \eta_-)
    (\psi_+ + \eta_+) \label{raw}
\eea
( Here the component derivatives ${\cal D}$ are covariant with respect
to both vector and twisted vector multiplets.)

This is followed by some
additional manipulations:  integration by parts, and the use of
relations that follow from the fact that $K$ is neutral under the separate
action of the two symmetries gauged by the vector multiplets, namely
\bea
\Phi K_{\Phi} - \bar{\Phi} K_{\bar{\Phi}} &=& 0   \non \\
\calX K_{\calX} -\bar{\calX} K_{\bar{\calX}} &=& 0  ~~~.
\eea
{} From these two equations, it is possible to derive many more by applying
various spinorial derivatives and evaluating at $\theta=0$.
Applying one derivative we obtain the following
superspace identities:
\bea
\Delp \Phi &=& (\Phib - \Phi) (\Delp \Phi + \Delp \calX)  \non\\
\Delm \Phi &=& (\Phib - \Phi) (\Delm \Phi + \Delm \bar{\calX})   \non\\
\Delpd \Phib &=& (\Phi - \Phib)(\Delpd \Phib + \Delpd \bar{\calX})
   \non\\
\Delmd \Phib &=& (\Phi - \Phib) (\Delmd \Phib + \Delmd \calX)   \non\\
\Delp \calX &=& (\bar{\calX}- \calX) (\Delp \Phi + \Delp \calX)  \non\\
\Delm \bar{\calX} &=& (\calX - \bar{\calX})
(\Delm \Phi + \Delm \bar{\calX})   \non\\
\Delpd \bar{\calX} &=& (\calX - \bar{\calX})
(\Delpd \Phib + \Delpd \bar{\calX})   \non\\
\Delmd \calX &=& (\bar{\calX} - \calX)(\Delmd \Phib
+ \Delmd \calX) \label{trick8}~~.
\eea
We stress
 once again that the derivatives of $K$ with respect to $\Phi, \Phib,
\bar{\calX}, \calX$ are implicit; for example, the first equation reads
\beq
K_{\Phi} \Delp \Phi = K_{\Phi \Phib} \Phib \Delp \Phi - K_{\Phi \Phi} \Phi
\Delp \Phi
+K_{\Phib {\cal X}} \Phib \Delp {\cal X} - K_{\Phi {\cal X}} \Phi \Delp {\cal
X}~~.
\eeq
The associated component expansions are:
\bea
\psi_+ &=& (\phib -\phi)(\psi_+ + \eta_+)  \non\\
\psi_- &=& (\phib -\phi)(\psi_- + \eta_-)  \non\\
\psi_\pd &=& (\phi -\phib)(\psi_{\pd} + \eta_{\pd}) \non \\
\psi_\md &=& (\phi -\phib)(\psi_\md + \eta_\md)  \non\\
\eta_+ &=& (\bar{\chi} -\chi)(\psi_+ + \eta_+)  \non\\
\eta_- &=& (\chi -\bar{\chi} )(\psi_- + \eta_-)  \non\\
\eta_\pd &=& (\chi -\bar{\chi})(\psi_\pd + \eta_\pd)  \non\\
\eta_\md &=& (\bar{\chi} -\chi)(\psi_\md + \eta_\md) .
\eea
Applying two derivatives gives us (in superspace) additional relations,
of which we need the following:
\bea
\Delm \Delpd \Phib &=& \Delpd \Phib(\Delm \Phi + \Delm \bar{\calX})
 + \Delm \Phi
 (\Delpd \Phib + \Delpd \bar{\calX})  \non \\
&&+ (\Phi - \Phib)[\Delm \Delpd \Phib + \Delm \Delpd \bar{\calX}
 - (\Delpd \Phib +
 \Delpd \bar{\calX})(\Delm \Phi + \Delm \bar{\calX})]  \non \\
\Delmd \Delp \Phi &=& \Delp \Phi(\Delmd \Phib + \Delmd \calX) + \Delmd \Phib
 (\Delp \Phi + \Delp \calX)  \non  \\
&&+ (\Phib - \Phi)[\Delmd \Delp \Phi + \Delmd \Delp \calX -(\Delp \Phi +
 \Delp \calX) (\Delmd \Phib + \Delmd \calX)] \non \\
\Delm \Delpd \bar{\calX}
 &=& \Delpd \bar{\calX}(\Delm \Phi + \Delm \bar{\calX}) + \Delm \calX
 (\Delpd \Phib + \Delpd \bar{\calX})   \\
&&+ (\calX - \bar{\calX})[\Delm \Delpd \Phib +
\Delm \Delpd \bar{\calX} - (\Delpd \Phib
 + \Delpd \bar{\calX})(\Delm \Phi + \Delm \bar{\calX})]  \non
\eea
The respective component expansions are:
\bea
-qW \phib &=& \psi_{\pd} (\psi_- +\eta_- ) +\psi_-( \psi_\pd + \eta_\pd )
\nonumber\\
&&+ (\f -\phib )[ -qW \ +i \bar{G} - (\psi_\pd +\eta_\pd )(\psi_- +\eta_-)]
 \non\\
-q \bar{W} \phi &=&  \psi_+(\psi_\md + \eta_\md)
+ \psi_\md(\psi_+ + \eta_+) \non \\
&&+(\phib - \phi)[-q \bar{W} \phi +iG-
(\psi_+ + \eta_+)(\psi_\md + \eta_\md)]  \non\\
i \bar{G} &=& \eta_\pd(\psi_- + \eta_-) + \eta_-(\psi_\pd + \eta_\pd) \non \\
&&+(\chi - \bar{\chi} )[-qW \phib +i\bar{G}
 - (\psi_\pd + \eta_\pd)(\psi_- + \eta_-)]  \label{compexp}
\eea

 Utilizing  \reff{raw} and  \reff{compexp},
one obtains the
following component lagrangian, which is symmetric with respect to all
the fields (again factors of $K$ should be reinstated as appropriate, so that,
e.g., a
term $PF \psi_+ \eta_-$ should be multiplied by $K_{\phi \phi
\tilde{\chi}}$)):
\bea
\lefteqn{{\cal S} = \int d^2 x  \{ -\half  { \calD}_{\m} \phi { \calD}_{\m}
\bar{\phi}
+\half  { \calD}_{\m}
\chi { \calD}_{\m} \bar{\chi} }\nonumber\\
&&-\half  \e_{\m \n} { \calD}_{\m}\phi{ \calD}_{\n}\chi +\half \e_{\m \n}
 {\calD}_{\m }  \bar{\chi} { \calD}_{\n}\bar{\phi} \nonumber\\
&&-\frac{i}{2} \psi_{\pd}   \Dmmarl \psi_+ +\frac{i}{2} \eta_{\pd}   \Dmmarl
\eta_+
-\frac{i}{2}  \psi_{\md} \Dpparl \psi_- +\frac{i}{2} \eta_{\md} \Dpparl \eta_-
\nonumber\\
&&+\frac{i}{2}  ( \psi_{\pd}\psi_+ +\eta_{\pd} \eta_+) {\cal D}_{\mm}
(\bar{\phi} - \phi
+\chi - \bar{\chi} ) \non \\
&&- i \psi_{\pd}\eta_+ {\cal D}_{\mm} (\phi + \bar{\chi} )
+i \eta_{\pd}\psi_+ {\cal D}_{\mm}(\bar{\phi} +\chi) \nonumber\\
&&+\frac{i}{2}  ( \psi_{\md}\psi_- +\eta_{\md} \eta_-) {\cal D}_{\pp}
(\bar{\phi} - \phi
-\chi + \bar{\chi}) \non \\
&&- i \psi_{\md}\eta_- {\cal D}_{\pp} (\phi + \chi)
+i \eta_{\md}\psi_- {\cal D}_{\pp}(\bar{\phi} +\bar{\chi} )\nonumber\\
&&+(\psi_{\md}+\eta_{\md})(\psi_{\pd}+\eta_{\pd})
(\psi_-+\eta_-)(\psi_++\eta_+)
\nonumber\\
&&-\bar{F}F +i\bar{F}(\psi_-+\eta_-) (\psi_++\eta_+)
+iF(\psi_{\md}+\eta_{\md})(\psi_{\pd}
+\eta_{\pd}) \nonumber\\
&&+  G\bar{G} +iG(\psi_{\pd}+\eta_{\pd})(\psi_-+\eta_-)
-i\bar{G}(\psi_{\md}+\eta_{\md})
(\psi_++\eta_+) \nonumber\\
&&+q \phi[ \l_- (\psi_{\pd}+\eta_{\pd}) -\l_+(\psi_{\md}+\eta_{\md})] \non\\
&& +q \bar{\phi}
[\l_{\md}(\psi_++\eta_+)- \l_{\pd}(\psi_-+\eta_-)] \nonumber\\
&&+qW[\psi_{\md}(\psi_++\eta_+) +
\bar{\phi}(\psi_{\md}+\eta_{\md})(\psi_++\eta_+)] \non\\
&&+q\bar{W}[\psi_-(\psi_{\pd}+\eta_{\pd}) -\phi
(\psi_{\pd}+\eta_{\pd})(\psi_-+\eta_-)]
\nonumber\\
&&+q'\bar{\chi} [\rho_-(\psi_++ \eta_+) - \rho_{\pd}(\psi_{\md}+\eta_{\md})]
\non\\
&&+q'
\chi [\rho_{\md}(\psi_{\pd}+ \eta_{\pd})- \rho_+(\psi_-+\eta_-)] \nonumber\\
&&+q'P[-\eta_+(\psi_-+\eta_-)+\chi (\psi_-+\eta_-)(\psi_++\eta_+)]  \non\\
&&+q'
\bar{P}
[-\eta_{\pd}(\psi_{\md}+\eta_{\md})+ \bar{\chi}
(\psi_{\md}+\eta_{\md})(\psi_{\pd}+\eta_{\pd})] \nonumber\\
&&+(q')^2 P\bar{P}\chi \bar{\chi} +i q'\bar{P}\bar{F} \bar{\chi} +iq'PF\chi
+\frac{i}{4}q' (H+\bar{H})(\chi +\bar{\chi} )\nonumber\\
&&-q^2W\bar{W}\phi \bar{\phi} +iqWG\bar{\phi}+iq \bar{W} \bar{G}\phi
+\frac{i}{4}q(J+\bar{J})(\phi +\bar{\phi}) \} ~.
\eea

Note that in the last  two lines the terms $J+\bar{J}$  and
$H+\bar{H}$ are
proportional to the  auxiliary fields of the vector multiplets.

In the above equations ${\cal D}$ is the component gauge covariant
derivative, and
\bea
{\cal D}_{\m}\phi {\cal D}_{\m} \bar{\phi} & =& {\calDpp}\phi {\calDmm}
\bar{\phi} +
{\calDmm}\phi {\calDpp} \bar{\phi}  \nonumber\\
\epsilon_{\m \n} {\cal D}_{\m}\phi {\calD}_{\n} \chi &=& {\calDpp}\phi
{\calDmm}
\chi -
{\calDmm}\phi {\calDpp} \chi ~~~.
\eea

We  consider now extensions of these results by including explicit
dependence on the field strengths of
vector and twisted vector multiplets.   First we note that the action in
\reff{KFX}
 can be extended to include
 multiplets of chiral and twisted chiral matter superfields,
$\Phi \to  \{\Phi^i \}$ and ${\cal X} \to \{ {\cal X}^{\rm I} \}$ (where $i =
1,...
,m$ and ${\rm I} = 1,...,n)$. The indices are just carried along passively
through the calculations. We take advantage of this by noting that
an action of the form
\beq
S=\int d^2 x \, d^4 \theta ~ K(\Phi , \Bar{\Phi}, {\cal X}, {\Bar{\cal X}};
{\cal P}, \, {\Bar {\cal P}}, \, {\cal W}, \, {\Bar {\cal W}} )
{}~~~ \label{KFPW}
\eeq
looks exactly like the case with $m = 2$ and $n = 2$, if we  replace
$\Phi  \rightarrow  \{ \Phi , \, {\cal P} \}$ and ${\cal X}  \rightarrow  \{
{\cal X},
\, {\cal W}  \}$. Arbitrary numbers of vector and twisted
vector multiplets can be included by simply re-interpreting the value
of the indices on $\Phi^i$ and ${\cal X}^{\rm I}$. For example,
the indices $i = 1,...,p$ are associated with chiral multiplets
while $i = p + 1,...,m$ are associated with twisted vector multiplets.
Similarly, the indices ${\rm I} = 1,...,q$ are associated with
twisted chiral multiplets while ${\rm I} = q + 1,...,n$ are associated with
vector multiplets.

As an example,  we give the general coupling of an
SU(2) vector multiplet to an SU(2) doublet of chiral scalars.  We have
$p = 2$, $m = 2$, $q = 0$ and $n = 3$.  We begin with \reff{KFPW} and for
simplicity
write only  the purely bosonic terms.
\bea
S &=& \int d^2 x \, d^4 \theta ~ K(\Phi , \Bar{\Phi}, 0, 0; 0
, \, 0, \, {\cal W}, \, {\Bar {\cal W}} ) |_{bosonic
{}~ fields} \nonumber\\
&=&\int d^2 x \, \Big\{~  -\half  K_{\phi^i {{\Bar \phi}}^j} { \calD}_{\m}
\phi^i { \calD}_{\m}  \Bar{\phi}^j +\half K_{W^{\rm I} {{\Bar W}}^{\rm J}}
{ \calD}_{\m} W^{\rm I} { \calD}_{\m} {\Bar W}^{\rm J} \nonumber\\
&&{~~~~~~~~~~~~}-\half  \e_{\m \n} K_{\phi^i W^{\rm J} } { \calD}_{\m}\phi^i
{ \calD}_{\n}W^{\rm J} +\half \e_{\m \n} K_{{\Bar W}^{\rm I} {\Bar \phi}^j }
 {\calD}_{\m } {\Bar W}^{\rm I} { \calD}_{\n}\bar{\phi}^j \nonumber\\
&&{~~~~~~~~~~~~}-K_{\phi^i {{\Bar \phi}}^j} F^i {\Bar F}^j + \frac 14 K_{W^{\rm
I} {{\Bar W}}^{\rm J}} [\, {\rm F}^{\rm I} (A) {\rm F}^{\rm J} (A) ~+~ {\rm
d}^{\rm I}
{\rm d}^{\rm J} \,]  \nonumber\\
&&{~~~~~~~~~~~~}-g^2  K_{{\phi}^i {{\Bar \phi}}^j} ([ W, \, \phi ] )^i ([
\Bar{W} , \, {\Bar \phi} ])^j \nonumber\\
&&{~~~~~~~~~~~~} +\ihalf  g K_{{\Bar \phi}^i W^{\rm J} }  ([ W , \, {\Bar
\phi}])^i [  {\rm d}^{\rm J} +i {\rm F}^{\rm J} (A)   ] \nonumber\\
&&{~~~~~~~~~~~~} + \ihalf g  K_{{\phi}^i {\Bar W}^{\rm J} } ([ \Bar{W}, \,
\phi ])^i [  {\rm d}^{\rm J} +i {\rm F}^{\rm J} (A)  ] \nonumber\\
&&{~~~~~~~~~~~~}+\frac{i}{4}g   K_{{\phi}^i } ([ (J+\bar{J}) , \, \phi ])^i
+\frac{i}{4}g   K_{{\Bar \phi}^i } ([ (J+\bar{J}) , \, {\Bar \phi} ])^i
{~} \Big\} ~~~. \label{KFW}
\eea
In this expression, it is understood that when a quantity such as $W$
appears without an index, we are referring to the corresponding
Lie-algebraic operator (i.e. $W \equiv W^{\rm I} t_{\rm I}$).

 By use of the mirror map operator, we can easily
obtain the mirror reflection, namely
\beq
\int d^2 x \, d^4 \theta ~ K(0 , 0, {\cal X}, \Bar{{\cal X}}; {\cal P}
, \, {\Bar {\cal P}}, \, 0, \, 0 ) = {\cal M}_m : \int d^2 x \, d^4
\theta ~ K(\Phi , \Bar{\Phi}, 0, 0; 0 , \, 0, \, {\cal W}, \,
{\Bar {\cal W}} )
\eeq
where at the component level we make the obvious replacement of fields.

As noted before, integrability conditions forbid minimal couplings of twisted
vector multiplets to  chiral scalar multiplets
 or  vector multiplets to  twisted chiral
scalar multiplets. Nonminimal couplings are possible
through Pauli-moment type terms.  This possibility
occurs because  vector multiplet field strengths are chiral or twisted
chiral  and can therefore be used in superpotential
actions.  The superpotentials
of  \reff{pot} can be generalized to include the respective vector
multiplets.
\bea
{\cal S}_c &=& \int d^2x d^2 \th U( \Phi , {\cal P}) + h.c. \non\\
 &=&\int d^2 x e^{-1} \lbrace ~ U_{\phi \, \phi}
 \psi_+  \psi_-  + U_{\phi  P} (
 \psi_+  \rho_-  + \rho_+ \psi_-  )
+  U_{ P P} \rho_+ \rho_-
{~~~~}\cr
&&{~~~~~~~~~~~~~~} -i  U_{\phi} F+
-\ihalf U_{P} [{\rm d}'+i{\rm F}(A')]   ~+~ h.c. \rbrace ,
\eea
and
\bea
{\cal S}_{tc}  &=& \int d^2x d \th^+ d \th^\md  \tilde{U}( {\cal X} , {\cal W})
+ h.c. \non\\
&=&\int d^2 x e^{-1} \lbrace ~ {\Tilde U}_{\chi
\chi} \eta_+ \eta_{\md} +
{\Tilde U}_{\chi W} ( \eta_+
\l_{\md}+ \l_+  \eta_{\md})   +
{\Tilde U}_{WW} \l_+\l_{\md}
{~~~~}\cr
&&{~~~~~~~~~~~~~~} -i
{\Tilde U}_{\chi } G -\ihalf {\Tilde U}_{W} [{\rm d}+i{\rm F}(A)]     ~~+~~h.c.
\rbrace~~.
\eea

\sect{Lagrangians for Matter Multiplets Coupled to
Supergravity}
In this section we consider matter-supergravity systems, and their
component actions. We concentrate primarily on the irreducible
$U_A(1)$ version of supergravity, but we present in subsection 7.5
results involving the nonminimal $U_V(1) \otimes U_A(1)$ theory.

\subsection{Chiral Multiplet}
In this subsection we obtain the component lagrangian for the kinetic term of a
 {\em covariantly} chiral
superfield  (defined by  $\Delb_\ad \F = \Dela \bar{\F} = 0$ )
coupled to the $U_A(1)$ version of $(2,2)$ supergravity.
The  chiral projection formula \reff{chproj}
\bea
{\cal S} _{\Phib \Phi} &=& \int d^2 x d^4 \th E^{-1} \bar{\Phi}\Phi  \non \\
&=& \int d^2 x e^{-1} [\Delsq + i\psi_{\mm}^{\md}  \Delp -i\psi_{\pp}^{\pd}
\Delm
+(   - \half \bar{B} - \psi_\pp^\md \psi_\mm^\pd + \psi_\mm^\md \psi_\pp^\pd )]
\Delbsq
 (\bar{\F}\F)|     \non \\
&=& \int d^2 x e^{-1}[(\Delsq \Delbsq \bar{\F}) \F| + (\Delp \Delbsq \bar{\F})
 (\Delm \F)|
 - (\Delm \Delbsq \bar{\F})(\Delp \F)| \non \\
&& + (\Delbsq \bar{\F}) (\Delsq \F)| + i\psi_{\mm}^{\md} \Delp (\Delbsq
\bar{\F} \F) |\non \\
&& -i \psi_{\pp}^{\pd} \Delm (\Delbsq \bar{\F} \F) |+ (  - \half \bar{B} -
\psi_\pp^\md \psi_\mm^\pd + \psi_\mm^\md \psi_\pp^\pd )(\Delbsq \bar{\F}) \F|]
{}~~.
\eea

We list the component expansions for the quantities
appearing above:
\bea
\Delp \Delbsq \bar{\F}| &=& i \bDpp \psi_\md - \psi_\pp^- \bDmm \bar{\f} -
\psi_\pp^-
\psi_\mm^\pd \psi_\pd - \psi_\pp^- \psi_\mm^\md \psi_\md +  \psi_\pp^\pd
\bar{F} \nonumber\\
\Delm \Delbsq \bar{\F}| &=& -i \bDmm \psi_\pd - \psi_\mm^+ \bDpp \bar{\f} +
\psi_\mm^+
\psi_\pp^\md \psi_\md + \psi_\mm^+ \psi_\pp^\pd \psi_\pd +  \psi_\mm^\md
\bar{F} \nonumber\\
\Delsq \Delbsq \bar{\F}| &=& \Delpp \Delmm \bar{\F}| + \half(\Delpd \Rb)\Delmd
\bar{\F}| + \half \Rb \Delbsq \bar{\F}| \nonumber\\
&=&\Delpp \Delmm \bar{\F}|
- [\bD_{[\pp} \psi_{\mm]}^{\pd} - \sihalf \psi_\mm^- \Rb|
+
 i \psi_{\pp}^+ \psi_\mm^{\pd} \psi_{\pp}^{\pd} + i \psi_{\pp}^{\md}
\psi_\mm^-
\psi_\mm^{\pd}
 + i \psi_{\pp}^- \psi_\mm^{\md} \psi_\mm^{\pd}]
\Delmd
\bar{\F}|  \nonumber\\
&&+ \half \Rb \Delbsq \bar{\F}| \nonumber\\
\Delpp \Delmm \bar{\F}| &=& \bDpp (\bDmm + \psi_\mm^\ad \Del_\ad) \bar{\F}|
-  \ihalf \psi_\pp^+ \Rb \Delmd \bar{\F} | \non  \\
&& + \psi_\pp^\pd [ \bDmm \Delpd + i \psi_\mm^+( \bDpp + \psi_\pp^\ad \Delad)
- \psi_\mm^\md \Delbsq ] \bar{\F} | \non \\
&& + \psi_\pp^\md [ \bDmm \Delmd + i \psi_\mm^-( \bDmm + \psi_\mm^\ad \Delad)
+ \psi_\mm^\pd \Delbsq ] \bar{\F}| \non \\
&=& \bDpp \bDmm \bar{\f} + \bDpp \psi_\mm^\pd \psi_\pd + \bDpp \psi_\mm^\md
\psi_\md
+ \psi_\pp^\pd \bDmm \psi_\pd + \psi_\pp^\md  \bDmm \psi_\md \non \\
&&-  \ihalf \psi_\pp^+ \bar{B} \psi_\md + i \psi_\pp^\pd \psi_\mm^+ \bDpp
\bar{\f}
+i \psi_\pp^\md \psi_\mm^- \bDmm \bar{\f} \non \\
&& + i \psi_\pp^\pd \psi_\mm^+ \psi_\pp^\md \psi_\md
+ i \psi_\pp^\md \psi_\mm^- \psi_\mm^\md \psi_\md
+ i \psi_\pp^\md \psi_\mm^- \psi_\mm^\pd \psi_\pd \non \\
&& -i \psi_\pp^\md \psi_\mm^\pd \bar{F} +i \psi_\pp^\pd \psi_\mm^\md \bar{F}
{}~~.
\eea
We have used the identities that appear in
the Appendix, as well as the component identifications
\bea
\F| = \f &,& \bar{\F}| = \bar{\f}  \non\\
\Delp \F | = \psi_+ &,& \Delpd \bar{\F}| = \psi_\pd  \non\\
\Delm \F | = \psi_- &,& \Delmd \bar{\F}| = \psi_\md  \non\\
\sihalf [\Delp , \Delm ]\F| = F &,& \sihalf [ \Delpd , \Delmd ]\bar{\F}|
= \bar{F}   ~~.
\eea

Once we have the complete component expansion,  we group
together similar terms (component bosonic, fermionic, and spinorial-bosonic).
For
each set,  once one
separates out the contribution from the extra gravitino pieces in the
connection as in \reff{conn} , the results are symmetric
in barred/unbarred quantities. To illustrate how this works, we collect
together all the bosonic terms, and get
\beq
-\bar{F} F + (\bDpp \bDmm \bar{\f}) \f + i (\psi_\pp^\md \psi_\mm^- +
\psi_\pp^- \psi_\mm^\md) (\bDmm \bar{\f}) \f ~~.
\eeq
We  separate out the gravitino terms in $\bDpp$ when it acts on
$\bDmm \bar{\f}$, as follows:
\bea
\bDpp (\bDmm \bar{\f}) &=& [e_\pp + \o_\pp M + \g_\pp \Mb] (\bDmm \bar{\f})\non
\\
&=& [e_\pp - \half (\o + \g)_\pp] \bDmm \bar{\f} \non \\
&=& [e_\pp - \vf_\pp] \bDmm \bar{\f} \non \\
&=& [e_\pp - \vp_\pp -i (\psi_\pp^\md \psi_\mm^- + \psi_\pp^- \psi_\mm^\md)
 (\bDmm \bar{\f})] \non \\
&=& \calDpp (\bDmm \bar{\f}) -i(\psi_\pp^\md \psi_\mm^- + \psi_\pp^-
\psi_\mm^\md)
 (\bDmm \bar{\f}) \label{grav} ~~,
\eea
where $\calDpp $ is the ordinary gravitational covariant derivative.

Substituting  into the bosonic terms above, we find that the gravitino
pieces cancel.
Furthermore, when  ${\bf D}$ acts on a scalar field, one can replace ${\bf
D}_\mm
\phib$ by $\pa_\mm \phib$, and inside the $d^2x e^{-1}$ integral we can
integrate by parts, leaving only $(-\bar{F} F - \di_\mm \bar{\f} \di_\pp \f)$.

  For the fermionic terms, the gravitino pieces from the connection combine
with similar terms that were produced explicitly by the component projections.
We also separate out the $U_A(1)$ connection $V_a'$ explicitly, as in the
following
example:
\bea
\bDpp \psi_\md &=& [e_\pp + \g_\pp \Mb] \psi_\md  \non \\
             &=& [e_\pp - \half  \g_\pp]\psi_\md  \non \\
          &=& [e_\pp - \half  \varphi_\pp -  \ihalf {V}_\pp '] \psi_\md  \non
\\
          &=& [e_\pp - \half  \vp_\pp - \ihalf {V }_\pp '
  -  \ihalf (\psi_\pp^\md \psi_\mm^- + \psi_\pp^- \psi_\mm^\md)]\psi_\md
\non \\
          &=& (\calDpp -  \ihalf {V }_\pp') \psi_\md
  - \ihalf (\psi_\pp^\md \psi_\mm^- + \psi_\pp^- \psi_\mm^\md)\psi_\md ~~.
\eea

   Summing all the terms, the final result for
the component lagrangian for the kinetic term of the chiral multiplet
coupled to the $U_A(1)$ version of (2,2) supergravity is
\bea
{\cal S}_{\Phib \Phi} &=&\int d^2 x d^4 \th E^{-1} \bar{\Phi}\Phi  \non \\
&=&\int d^2 x  ~e^{-1} \{-\bar{F}F - \pa_\mm \bar{\phi} \pa_\pp \phi
+ i (\calDpp \psi_{\md})\psi_-
+ i (\calDmm \psi_{\pd})\psi_+ \non \\
&&~~~~~~ + \half {V }_{\pp}'\psi_\md \psi_- - \half
   {V }_\mm'  \psi_{\pd}\psi_+ \non \\
&&~~~~~~ - [\pa_{\pp}\phi + \half \psi_{\pp}^{\a} \psi_{\a}] \psi_\mm^{\pd}
\psi_{\pd}
-[\pa_\mm \phi + \half \psi_\mm^{\a} \psi_{\a}] \psi_\pp^\md \psi_\md \non \\
&&~~~~~~ - [\pa_\pp \bar{\phi}
+ \half \psi_{\pp}^\ad \psi_\ad] \psi_\mm^+ \psi_+ - [\pa_\mm \bar{\phi}
+ \half \psi_\mm^\ad \psi_\ad] \psi_\pp^- \psi_- \}  \label{chiral}
\eea
which agrees with the expression first derived in ref. \cite{BS}.

\subsection{Twisted Chiral Multiplet}
To obtain the component lagrangian for the kinetic term of a
twisted chiral multiplet, we start with the same {\it chiral} projection
formula as above, except that now ${\cal X}$ is subject to the twisted
chirality
conditions $\Delpd {\cal X} =
\Delm {\cal X} = \Delp \bar{{\cal X}} = \Delmd \bar{{\cal X}} = 0$.
\bea
{\cal S}_{ \bar{\cal X} {\cal X}} &=&- \int d^2 x d^4 \th E^{-1} \bar{\cal
X}{\cal X}  \non \\
&=&- \int d^2 x e^{-1} [\Delsq + i\psi_{\mm}^{\md}  \Delp -i\psi_{\pp}^{\pd}
\Delm
+(   - \half \bar{B} - \psi_\pp^\md \psi_\mm^\pd + \psi_\mm^\md \psi_\pp^\pd )]
\Delbsq
 (\bar{\cal X}{\cal X})|  \non \\
&=& -\int d^2x e^{-1} \{ -i (\Delpp \Delm \bar{{\cal
X}})(\Delmd {\cal X})|
+ (\Delm \Delpd \bar{{\cal X}})(\Delp \Delmd {\cal X})| \non \\
&&~~~~~~ + (\Delpp \bar{{\cal
X}})(\Delmm {\cal X}) |
+ i (\Delpd \bar{{\cal X}})(\Delp \Delmm {\cal X})|  \non\\
&&~~~~~~+i \psi _\mm^\md[  i(\Delpp \bar{{\cal X}})(\Delmd
{\cal X})| - (\Delpd
\bar{{\cal X}}) (\Delp \Delmd {\cal X})| ]  \non\\
&&~~~~~~-i \psi_\pp^\pd [  -i(\Delmm {\cal X})(\Delpd
\bar{{\cal X}})| + (\Delmd
{\cal X}) (\Delm \Delpd \bar{{\cal X}})|]   \non\\
&&~~~~~~+(   - \half \bar{B} - \psi_\pp^\md \psi_\mm^\pd + \psi_\mm^\md
 \psi_\pp^\pd )
 (\Delpd \bar{{\cal X}})(\Delmd {\cal X})| \}
\eea
The components of the twisted chiral multiplet are defined by covariant
projection:
\bea
{\cal X}| = \chi&,& \bar{{\cal X}}| = \bar{\chi}  \non\\
\Delp {\cal X} | = \eta _+ &,& \Delpd \bar{{\cal X}}| = {{\eta }}_\pd  \non\\
\Delmd {\cal X}| = \eta _\md &,& \Delm \bar{{\cal X}} | = {{\eta }}_-  \non\\
\sihalf [\Delp , \Delmd ] {\cal X}|
= G &,& \sihalf [\Delpd ,\Delm ] \bar{{\cal X}}| = \bar{G}  ~~.
\eea
In this case, the required component derivative expansions are:
\bea
\Delpp \Delm \bar{{\cal X}}| &=& \bDpp {{\eta }}_- -i \psi_\pp^\pd \bar{G} + i
\psi_\pp^\md(\bDmm \bar{\chi} + \psi_\mm^- {{\eta }}_- + \psi_\mm^\pd {
{\eta }}_\pd)  \non\\
\Delp \Delmm {\cal X}| &=& - \sihalf \bar{B} \eta _\md + \bDmm \eta _+ + i
\psi_\mm^\pd
(\bDpp \chi + \psi_\pp^+ \eta _+ + \psi_\pp^\md \eta _\md) +i \psi_\mm^\md G
\non\\
\Delpp \bar{{\cal X}}| &=& \bDpp \bar{\chi} + \psi_\pp^\pd {{\eta }}_\pd +
\psi_\pp^- {{\eta }}_-
\non\\
\Delmm {\cal X}| &=& \bDmm \chi + \psi_\mm^+ \eta _+ + \psi_\mm^\md \eta _\md
{}~~.
\eea
We follow the same procedure as in the case of the chiral multiplet.
Collecting all the terms in the sum
and separating out the gravitino pieces as before, we obtain
\bea
{\cal S}_{ \bar{\cal X} {\cal X}} &=&-\int d^2 x d^4 \th E^{-1} \bar{{\cal X}}
{\cal X}  \non \\
&=&  \int d^2 x  ~e^{-1} \{ -\bar{G}G - \pa_\mm \bar{\chi} \pa_\pp \chi
+ i (\calDpp {{\eta }}_-) \eta _{\md}
+i (\calDmm \eta _+) {{\eta }}_\pd \non\\
&&~~~~~~ + \half {V }_{\pp}'\eta _\md {{\eta }}_-
- \half {V }_\mm' {{\eta }}_\pd \eta _+ \non \\
&&~~~~~~ - [\pa_{\pp}\chi + \half (\psi_{\pp}^\md \eta _\md + \psi_\pp^+ \eta
_+)]
   \psi_\mm^{\pd} {{\eta }}_{\pd}\non \\
&&~~~~~~-[\pa_\mm \chi + \half (\psi_\mm^\pd {{\eta }}_\pd + \psi_\mm^-
  {{\eta }}_-)] \psi_\pp^\md \eta _\md \non \\
&&~~~~~~- [\pa_\pp \bar{\chi}
+\half (\psi_{\pp}^- {{\eta }}_- + \psi_\pp^\pd {{\eta }}_\pd)]
 \psi_\mm^+\eta _+ \non \\
&&~~~~~~  - [\pa_\mm \bar{\chi}
+ \half (\psi_\mm^+ \eta_+ + \psi_\mm^\md \eta_\md)] \psi_\pp^- {{\eta}}_- \}
\label{twist}
\eea
as the component lagrangian for the kinetic term of the twisted chiral
multiplet
coupled to the $U_A(1)$ version of (2,2) supergravity.

Note that if we interchange $\{-\}$ and $\{\md\}$ in the \reff{twist} above
(and change the
$U(1)$ charge of the left-moving spinors),
we get exactly the result \reff{chiral} for the ordinary chiral multiplet.
We also point out that both \reff{chiral} and \reff{twist} are valid for the
$U_A(1)$ version of supergravity, but that to get the corresponding
expressions for the $U_V(1)$ version, one just interchanges $\{-\}$ and
$\{\md\}$.

\subsection{Potentials and Scalar-Vector Models}

To begin, let us calculate the component form of the potential terms
in \reff{pot} but in the presence of a $U_A (1)$ supergravity background.
We apply the chiral and twisted
chiral density projectors of  \reff{chproj} and \reff{twist density}  to
$U(\Phi )$ and
 $\tilde{U}( {\cal X})$ respectively and obtain the component actions
\bea
{\cal S}_c  &=&\int d^2 x e^{-1} \lbrace ~ U'' (\phi ) \psi_+ \psi_- ~+~
U' (\phi ) [-i F + i \psi_\mm^{\md} \psi_+ - i \psi_{\pp}^{\pd}
\psi_+  ] \cr
&&{~~~~~~~~~~~~~~} -~ U (\phi )  [  \half \bar{B} ~+~ \psi_{\pp}^\md
\psi _\mm^{\pd} ~-~ \psi_\mm ^\md \psi_{\pp}^{\pd} \, ] ~ ~+~~h.c.\rbrace ~~~,
\label{UC}
\eea
for the chiral case and
\bea
{\cal S}_{tc} &=& - \int d^2 x e^{-1} \lbrace ~ {\Tilde U}'' (\chi ) \eta_\md
\eta_+~+~  {\Tilde U}' (\chi ) [i G ~+~ i \psi_\pp^{\pd} \eta_{\md} ~-~
i \psi_{\mm}^- \eta_+ \, ] \cr
&&{~~~~~~~~~~~~~~} +~ {\Tilde U} (\chi )  [ \, \psi_{\pp}^{\pd}
\psi_{\mm}^{-} ~+~ \psi_{\pp}^{-}\psi_{\mm}^{\pd}) \, ]~~+~ ~h.c. \rbrace ~~~,
\label{UTC}
\eea
for the twisted chiral case.

We  can discuss now  local actions for the
chiral and twisted chiral multiplets with potential terms,
given by the sum of  \reff{chiral}, \reff{twist}, \reff{UC} and \reff{UTC}.  In
the following, we
are assuming that the potentials have the dimensions of mass.
We assume flat kinetic terms for the two types of multiplets:
$ \int d^2 x d^4 \q  E^{-1}{\Bar \Phi} \Phi$ for the chiral multiplet and
$-\, \int d^2 x d^4 \q  E^{-1} {\Bar {\cal X}} {\cal X}$  the twisted chiral
multiplet.  We obtain
\bea
  {\cal S} &=& -\int d^2 x  ~e^{-1} \lbrace ~ \pa_\mm \bar{\phi} \pa_\pp \phi
{}~+~
\pa_\mm \Bar{\chi} \pa_\pp \chi ~+~ \bar{F}F ~+~  \Bar{G}G \cr
&&{~~~~~~~~~~~~~~~~~~~}\, +~ i (\calDpp \psi_{\md})\psi_-
+ i (\calDmm \psi_{\pd})\psi_+ ~+~ \half {V}_{\pp}'\psi_\md \psi_- - \half
  V_\mm' \psi_{\pd}\psi_+ {~~~~~~~~~~} \cr
&&{~~~~~~~~~~~~~~~~~~~} +~ i (\calDpp {{\eta}}_-) \eta_{\md} ~+~ i
(\calDmm \eta_+) {{\eta}}_\pd + \half V_{\pp}'\eta_\md {{\eta
}}_- - \half V_\mm'{{\eta}}_\pd \eta_+ \non \\
&&{~~~~~~~~~~~~~~~~~~~} - [\pa_{\pp}\phi + \half \psi_{\pp}^{\a} \psi_{\a}]
\psi_\mm^{\pd}  \psi_{\pd} -[\pa_\mm \phi + \half \psi_\mm^{\a} \psi_{\a}]
\psi_\pp^\md \psi_\md {~~~} \cr
&&{~~~~~~~~~~~~~~~~~~~} - [\pa_\pp \bar{\phi}
+ \half  \psi_{\pp}^\ad \psi_\ad] \psi_\mm^+ \psi_+ - [\pa_\mm \bar{\phi}
+ \half \psi_\mm^\ad \psi_\ad] \psi_\pp^- \psi_- {~~~} \cr
&&{~~~~~~~~~~~~~~~~~~~} -  [\pa_{\pp}\chi + \half (\psi_{\pp}^\md \eta_\md
 + \psi_\pp^+ \eta_+)]
   \psi_\mm^{\pd} {{\eta}}_{\pd}  \cr
&&{~~~~~~~~~~~~~~~~~~~} -  [\pa_\mm \chi + \half (\psi_\mm^\pd {{\eta}}_\pd
 + \psi_\mm^-
  {{\eta}}_-)] \psi_\pp^\md \eta_\md  \cr
&&{~~~~~~~~~~~~~~~~~~~} -  [\pa_\pp \Bar{\chi}
+\half (\psi_{\pp}^- {{\eta}}_- + \psi_\pp^\pd {{\eta}}_\pd)]
 \psi_\mm^+\eta_+   \cr
&&{~~~~~~~~~~~~~~~~~~~} -  [\pa_\mm \Bar{\chi}
+ \half (\psi_\mm^+ \eta_+ + \psi_\mm^\md \eta_\md)] \psi_\pp^- {{\eta}}_-
 \cr
&&{~~~~~~~~~~~~~~~~~~~~~~} ~+~ [~ U'' (\phi ) \psi_+ \psi_- ~+~
U' (\phi ) (-i F ~+~ i \psi_\mm^\md \psi_+ ~-~ i \psi_{\pp}^{\pd}
\psi_+ \, ) \cr
&&{~~~~~~~~~~~~~~~~~~~~~~}~ -~ U (\phi )  ( \half \bar{B} ~+~ \psi_{\pp}^{\md}
\psi_\mm^{\pd} ~-~ \psi_\mm^{\md} \psi_{\pp}^{\pd}  )   \cr
&&{~~~~~~~~~~~~~~~~~~~~~~} ~+~  {\Tilde U}'' (\chi ) \eta_\md \eta_{
+} ~-~  {\Tilde U}' (\chi ) (i G ~+~ i \psi_\pp^{\pd} \eta_{\md} ~-~ i
\psi_{\mm}^{-} \eta_+ )  \cr
&&{~~~~~~~~~~~~~~~~~~~~~~}~ -~ {\Tilde U} (\chi )  (  \psi_{\pp}^{\pd}
\psi_{\mm}^{-} ~+~ \psi_{\pp}^{-}\psi_{\mm}^{\pd}  ) ~+
{\rm h.} {\rm c.} ~]  ~ \rbrace ~~. \label{acti}
\eea

This expression can be generalized trivially to any number of chiral and
twisted chiral multiplets. In particular, if we set the potential terms to zero
and
consider the case of just two multiplets, one chiral and one twisted chiral,
two chiral, or two twisted chiral, the action above  describes the mirror
symmetric,
or the two types of nonsymmetric, $N=2$ superstrings, whose possible existence
was
mentioned in \cite{GLO} some time ago.\footnote{
Recently, we have found evidence that the action for the N = 4 superstring
possesses this same \newline ${~~~~~}$ type of ambiquity. There appear to be
 {\em four} distinct N = 4 superstring actions and
\newline ${~~~~~}$ six distinct mirror map operators.}

Outside  the context of $N = 2$ superstrings, we can think of the potential
terms above as massive perturbations of $N = 2$ superstring theory. We note
that the potentials introduce mass for the spinors in the two multiplets
in very different ways. The spinor in the chiral multiplet   has a
Majorana-type mass term while the spinor in the twisted chiral multiplet has
a Dirac-type mass term.

If we  set all the fermions above to zero we obtain
\bea
 &-\, \int d^2 x  ~e^{-1} \lbrace ~ \pa_\mm \bar{\phi} \pa_\pp \phi ~+~ \pa_\mm
\Bar{\chi} \pa_\pp \chi ~+~ \bar{F}F ~+~  \Bar{G}G \,{~~~~~~~~~~~~~~~~~~~~~~}
\cr
&{~~~~~~~~~~~~~}  +~ [i U' (\phi )  F  +~ \half U (\phi )  \bar{B}
 ~+~  i{\Tilde U}' (\chi )  G ~+~ {\rm h.} {\rm c.}  ~ ] ~ \rbrace ~~~. \, \,
\eea
After eliminating the $F$ and $G$ auxiliary fields by their equations of
motion, potentials
are generated for the scalar fields. However, the presence of the supergravity
auxiliary field $B$ introduces an asymmetry in the constraints on
the potentials. The $B$ auxiliary field imposes the condition $U(\phi) = 0$.
 The trace of the zweibein
also imposes the conditions $U' (\phi)  = 0$ and  ${\Tilde U}' (\chi) = 0$.
If we think of these conditions as defining hypersurfaces in the $\phi$-space
and in the $\chi$-space, the two classes of surfaces are different, since one
can have
a  constant solution  ${\Tilde U} = {\Tilde U}_0$  whose  effect is to
produce a type of mass term for the gravitino. This
particular deformation of the mirror symmetric $N = 2$ superstring may
possess some interesting properties.

We next  consider scalar-vector matter systems coupled
to $2D$ supergravity. This can be done by assuming that in the
generalization of the
action \reff{acti} to one with  several scalar multiplets $\Phi^i$, ${\cal
X}^I$
 some of them are replaced by
 vector multiplets ${\cal P}$, ${\cal W}$.
The main effect of this is that, aside from the replacement of physical
components (e.g.
matter spinor into gauge spinor, etc.), one matter auxiliary field is replaced
by
a component (supercovariantized) gauge field strength (similar to
the discussion in section 3).
For example, substituting a twisted vector multiplet ${\cal P}$ for the chiral
multiplet
$\Phi'$ in the superpotential $U(\Phi, \Phib')$ leads to the component
expression
\bea
{\cal S}_c ~=~&& \int d^2x d^2 \th U( \Phi , {\cal P}) + h.c. \non\\
 =&&\int d^2 x e^{-1} \lbrace ~ U_{\phi \phi }
 \psi_+   \psi_-   ~+~ U_{\phi  P } ( \,
 \psi_+  \rho_-   ~+~ \rho_+  \psi_-    \,)
+~  U_{ P    P } \rho_+  \rho_-
{~~~~}\cr
&&{~~~~~~~~~~~~~~} +  U_{\phi } [\, -iF ~+~ i \psi_\mm^\md \psi_+
-~ i \psi_{\pp}^{\pd} \psi_+  \, ] \cr
&&{~~~~~~~~~~~~~~} -\ihalf U_{P } [ {\rm d}' + i {\rm F}(A') -\half (\bar{B} P
+B \bar{P})
 ] \cr
&&{~~~~~~~~~~~~~~} -~ U (\phi , \, P )  [ \, \half \Bar{B} ~+~ \psi_{\pp}^\md
 \psi_\mm^{\pd} ~-~ \psi_\mm^\md \psi_{\pp}^{\pd} \, ] ~ ~+~ h.c. \rbrace ~~~,
\eea
and  in a similar manner,
\bea
{\cal S}_{tc} ~=~&& \int d^2x d \th^+ d \th^\md  \tilde{U}( {\cal X} , {\cal
W}) + h.c. \non\\
= &&\int d^2 x e^{-1} \lbrace ~ {\Tilde U}_{\chi
\, \chi } \eta_\md \eta_ +  ~+~
{\Tilde U}_{\chi  \, W } (\, \eta_\md
\l_ +   ~+~ \l_\md   \eta_ +  \,)   ~+~
{\Tilde U}_{W  \, W } \l_ \md  \l_ +
{~~~~}\cr
&&{~~~~~~~~~~~~~~} +~
{\Tilde U}_{\chi  } [\, iG  ~+~ i \psi_\pp^{\pd} \eta_{\md}
  ~-~ i \psi_{\mm}^{-} \eta_{+}  \, ]
  ~+~ \ihalf {\Tilde U}_{W  } \,{\rm F} (A)   \cr
&&{~~~~~~~~~~~~~~} +~ {\Tilde U} (\chi , \, W )  [ \, \psi_{\pp}^{\pd}
\psi_{\mm}^{-} ~-~ \psi_{\pp}^{-}\psi_{\mm}^{\pd}) \, ] ~~+~ h.c. \rbrace ~~~.
\eea

Here we should mention an important difference between a
vector multiplet and a twisted vector multiplet, when coupled to $U_A(1)$
supergravity.  The component  spin-1 field strength of the twisted vector
multiplet (in addition to containing supercovariantized derivatives
of the component gauge field) is uniformly
modified in the Lagrangian above according to the replacement rule
(which follows from the commutator algebra of (3.1) in the presence
of supergravity)
\beq
{\rm F} (A') \to {\rm F}(A') ~-~ \frac 12 [ {\Bar B} P
  ~+~ {B} {\Bar P}]
\eeq
This replacement rule does not occur for a vector multiplet.
The appearance of the supergravity auxiliary field has the following
effect.   From the kinetic action $\int d^2 x  d^4 \th \Bar{\cal P} {\cal P}$
one
generates the component terms $\frac{1}{4} [{\rm F}(A') - \half (\bar{B}P + B
\Bar{P})]^2$.
As has been noted  in the $N = 1$ case \cite{glass},
  in the absence of a potential for the chiral matter
superfields, the supergravity auxiliary field's equation of motion forces
trivial dynamics for the twisted vector multiplet.

\subsection{Models with Gauged $U_V (1)$ or $U_A (1)$}

In a $(2,2)$ string-type theory the
matter multiplets are neutral under the tangent space $U_A(1)$ or $U_V(1)$
transformations.
One  way in which a superconformal $(2,2)$ model may be deformed is for
the $U_V (1)$ or $U_A (1)$ charge to be realized in a non-trivial
manner on the scalar multiplets  coupled to supergravity.
We  investigate  here the
resulting structure.

We  fix the form of the supergravity multiplet to correspond
to the $U_A (1)$ version of supergravity. For the matter scalar
superfields, we choose some number $N_C$
of chiral multiplets and some number $N_T$ of twisted chiral
multiplets.\footnote{For the N = 2 superstring we only have the choices
$N_C = 2$, $N_T = 0$,  $N_C = 1$, $N_T = 1$, and \newline ${~~~~~}$ $N_C = 0$,
$N_T = 2$.  We may call these the $C^2$, $CT$ and $T^2$ N = 2 superstrings,
respectively.}  For the sake of simplicity, we choose only flat kinetic terms.
However, we want  the matter
multiplets to carry a non-trivial realization of the charge that is
gauged by the vector in the supergravity multiplet. Since we
are using the $U_A (1)$ form of supergravity, the integrability condition
requires the chiral multiplets to  be neutral
under the $U_A(1)$ charge generator  while the twisted chiral
multiplet  transforms non-trivially:
\beq
 [ ~ {\cal Y}' \, , \, \Phi ~] ~=~ 0 ~~~~,~~~~
[ ~ {\cal Y}' \, , \, {\cal X} ~] ~=~ i {\cal Q}' \, {\cal X} ~~~,
\eeq
where ${\cal Q}'$ is an arbitrary real $N_T \times N_T$ diagonal matrix.
Since the chiral fields do not carry the $U_A(1)$ charge, their coupling
to supergravity is unchanged and their lagrangian given by \reff{twist}. For
the twisted
chiral superfields we have first
\beq
- \, \int d^2 \s d^4 \q E^{-1} ~{\Bar {\cal X}} \, {\cal X} ~=~ - \,
\int d^2 \s d^2 \q {\cal E}^{-1} ~ [\,  (\Del_{\pd} {\Bar {\cal X}})
(\Del_{\md} {\cal X} ) ~-~ \frac 12 \, R (\, {\Bar {\cal X}} {\cal Q}'
{\cal X} \,) ~ ] ~~~.
\eeq
Applying the chiral projection formula in \reff{chproj}, we obtain
\bea
\lefteqn{ {\cal S}_{{\cal Q}'} = \int d^2 x \, e^{-1} \Big\{
 \frac 12 [\,
\Del_- R \, - \, i \psi_{\mm}^{\md} R \, ] {\Bar \chi } {\cal Q}' \eta_+
 ~-~ \frac 12 [\, \Del_+ R \, - \, i \psi_{\pp}^{\pd} R \, ] \eta_- {\cal
Q}' \chi}  \non \\
&&- \frac 12 [\,
\Del_\md \Rb \, - \, i \psi_{\mm}^- \Rb \, ] \eta_\pd {\cal Q}' \chi
 ~+~ \frac 12 [\, \Del_\pd \Rb \, - \, i \psi_{\pp}^+ \Rb \, ] \eta_\md {\cal
Q}' \Bar{\chi}  \non \\
&&-  \frac 12 [\, {\Bar {\cal X}} {\cal Q}' {\cal X} \,
 ] \Delta ~ - ~ \frac 14  R {\Bar R} [\, {\Bar {\cal X}} ({\cal Q}')^2
{\cal X} \, ]  ~+~ \frac 12 R (\, \eta_- {\cal Q}' \eta_+ \, ) ~ \Big\} \Big|
 \label{SQ}
\eea
where
\beq
\Delta \equiv [~ \Del_+ \Del_- R \, + \,  i \psi_{\mm}^{\md} \Del_+ R
\, - \,  i \psi_{\pp}^{\pd} \Del_- R \, - \, R ( \frac 12 {\Bar R}
\, + \, \psi^\md_{\pp} \psi^{\pd}_{\mm} \,- \, \psi^\md_{\mm} \psi^{\pd}_{
\pp} ) ~ ] ~~~~.
\eeq

It is clear that  (super)conformal invariance is broken because this result
depends explicitly  upon the
supergravity auxiliary field $B$.  This breaking is seen most
explicitly if we retain only the bosonic terms in  ${\cal S}_{{\cal Q}'}
$ and note that the term involving $\Delta$ takes the form
\beq
\frac 12  [\, {\Bar {\cal X}} {\cal Q}' {\cal X} \, ]\D  ~~\to ~~
\frac 12 {\cal R} [\, {\Bar {\cal X}} {\cal Q}' {\cal X} \, ] ~~~.
\eeq
where ${\cal R}$ is the $2D$ curvature scalar. In $4D$, terms of this form lead
to an ``improvement'' of the spin-0 energy-momentum tensor. For a special
nonvanishing choice of ${\cal Q}'$ the $4D$ coupling is conformal. In $2D$,
precisely the opposite happens. For all nonvanishing ${\cal Q}'$ such a term
``degrades'' the conformal coupling of a free spin-0 field coupling
to $2D$ gravity.

In fact, this is an illustration of a point not generally recognized.
  We state this result in the form of a theorem:

${~~~~~~}$ {\it {In 2D, if the spin-0 field of a matter supermultiplet carries
a non-trivial \newline ${~~~~~~}$ realization of an internal symmetry charge
that is gauged by a spin-1 field \newline ${~~~~~~}$  in the superconformal
multiplet, the action for the spin-0 field is neither
\newline ${~~~~~~}$ conformally nor superconformally invariant.}}

\noindent If ${\cal Q}'$ is small, ${\cal S}_{\cal Q}'$ may be regarded as a
massless perturbation of the $CT$ or $T^2~ N = 2$ superstring.

Before we leave the result in \reff{SQ}, it is worth mentioning one of its
consequences for the pure gauge formulation of $N = 2~ U_A(1)$ supergravity.
There is a form of $U_A(1)$ supergravity where the usual auxiliary
scalar fields are replaced by the supercovariantized field strengths of a
complex
central  charge generator ${\cal Z}$. This form of $2D$, $N = 2$ supergravity
may be regarded as the result of a dimensional reduction of one of the
original off-shell forms of $4D$, $N = 1$ supergravity \cite{SW}.  The
commutator algebra of this formulation
is given by
\beq
 [\, \nabla_+ ~,~ \nabla_+ \} ~=~ 0 ~~~,~~~ [\, \nabla_- ~,~ \nabla_- \} ~=~ 0
 ~~~, ~~~ [\, \nabla_+ ~,~ {\nabla_{\md}} \} ~=~ 0 ~~~,
$$
$$
[\, \nabla_+ ~,~ { \nabla}_- \} ~=~ 2 \, { {\cal Z}} ~-~ \Rb \Mb ~~~, ~~~
[\, \nabla_{\pd} ~,~ {\nabla}_{\md} \} ~=~ - 2 \, {\Bar {\cal Z}} ~+~ R M ~~~,
$$
$$ [\, \nabla_+ ~,~ {\nabla}_{\pd} \} ~=~ i  \nabla_{\pp} ~~~,~~~
[\, \nabla_- ~,~ {\nabla}_{\md} \} ~=~ i  \nabla_{\mm} ~~~ ,
\eeq
where ${\cal Z}$ annihilates all of the
supergravity superfields, $[{\cal Z}, \Del_A \} = 0$.
All higher dimension commutators remain unchanged with the
exception of
\bea
[\Delpp, \Delmm ] &=&  \half (\Delp R) \Delm +\half (\Delm R) \Delp
        - \half (\Delpd \Rb) \Delmd - \half (\Delmd \Rb) \Delpd \non \\
&& -  \half R \Rb \Mb - \half R \Rb M + (\Delsq R) M -(\Delbsq \Rb) \Mb
\non \\
&&  ~+~ R \, {\cal Z} ~+~ {\Bar R} \, {\Bar {\cal Z}} ~~~~.
\eea

  The main point
of these equations is that this modified form of $2D$,
$ N = 2$ supergravity at the
component level amounts to the substitution for the auxiliary field
\beq
B ~~\to ~~ {\rm F}(U + i V) = \pa_{\pp} (U ~+~ i V)_{\mm} -  \pa_{\mm}
(U ~+~ i V)_{\pp} ~-~ 2 [~ \psi_{\pp}^\md \psi_\mm^{\pd}
{}~-~ \psi_\mm^\md \psi_{\pp}^{\pd} ~] ~~~,
\eeq
in terms of new real gauge fields $(U_{\pp}, U_{\mm})$ and $(V_{\pp},
V_{\mm})$.
The result of \reff{SQ} shows that in  the presence of a matter multiplet that
carries the $U_A(1)$ charge, the gauge fields $U_a$ and $V_a$ appear in the
action with the standard form of their kinetic terms multiplied
by matter scalars.
\beq
\frac 14  R {\Bar R} [\, {\Bar {\cal X}} ({\cal Q}')^2
{\cal X} \, ] ~~\to ~~ \frac 14 [\,  {\rm F}^2 (U) ~+~ {\rm F}^2(V) \, ] ~
[\, {\Bar {\cal X}} ({\cal Q}')^2
{\cal X} \, ] ~~~~.
\eeq
The supergravity auxiliary fields of
this formulation become physical degrees of freedom with
the vacuum value of the operator $[\, {\Bar {\cal X}} ({\cal Q}')^2
{\cal X} \, ]$ setting the scale of the central charge coupling
constants.  The replacement of the auxiliary fields by supercovariant field
strengths occurs uniformly in this modified form of $U_A(1)$ supergravity.
In particular, it occurs in the chiral density projector. Under this
circumstance, the quantity $\int d^2 x d \theta^+ d \theta^- {\cal E}^{-1}$
becomes a topological quantity proportional to a $U(1) \otimes U(1)$ index.
(In the standard form of the chiral projector,
after the replacement of the complex auxiliary field by the
complex supercovariant spin-1 field strength, the quadratic gravitini terms
in the non-derivative part of the density projector cancel.)

\subsection{ Classical Mirror Symmetric Models}

The existence of the mirror map in  $2D$, $N=2$ models is a simple consequence
of
the presence of a $U(1)$ charge in the supersymmetry algebra. It leads to a
pairing
 of  nonlinear $\s$-models represented by the action
\bea
{\cal S}_{\s} &=& \int d^2 x\, d^2 \th \, d^2 {\Bar \th} ~
K(\Phi, \, {\Bar \Phi} ; {\cal X} ,\, {\Bar {\cal X}}) \cr
&~& ~~~~ +~ \Big[ ~ \int d^2 x \, d^2 \th  ~ U(\Phi)  + {\rm {h. \, c.}}
{}~ \Big]  \cr
&~& ~~~~ +~ \Big[ ~ \int d^2 x \, d \th^+ \, d \th^{\md} ~ {\Hat U}({\cal X})
+ {\rm {h. \, c.}} ~ \Big]  ~~~~. {~~~~~~~~}
\eea
and the action
\bea
{\cal S}_{\tilde{\s}} &=& \int d^2 x \, d^2 \th \, d^2 {\Bar \th} ~
[\, - K( {\cal X} , \, {\Bar {\cal X}} ; \Phi, \, {\Bar \Phi}) \,] \cr
&~& ~~~~ +~ \Big[ ~\int d^2 x \, d \th^+ \, d \th^{\md} ~ U({\cal X}) +
{\rm {h. \, c.}} ~ \Big]  \cr
&~& ~~~~ +~ \Big[ ~ \int d^2 x \, d^2 \th ~ {\Hat U}(\Phi) + {\rm {h. \,
c.}} ~ \Big]  ~~~~. {~~~~~~~~}
\eea
with  $K$  real and $U$ and ${\Hat U}$  holomorphic.  It is generally  believed
that this
pairing is related to the pairing of mirror manifolds in Calabi-Yau
compactifications, but to
the best of our knowledge the precise connection has not been established.

In the remainder of this subsection we concentrate on the class
of models which  possess mirror symmetry, i.e. are  actually invariant
 under the map that interchanges $\th^- $ and $\th^\md$.
For rigid $N=2$ models the issue is very simple. As discussed in section 2,
the lagrangians corresponding to  $D$-terms must
be odd under the interchange of chiral and twisted chiral scalar or vector
multiplets,
while $F$-terms and twisted $F$-terms must go into each other under the map.
We discuss now the coupling of mirror-symmetric rigid models to supergravity.
Obviously, in order to maintain the symmetry, we must couple to
 the $U_V(1) \otimes U_A(1)$ version.

We note that using the reducible version of supergravity is perfectly natural
if
one regards the $2D$, $N=2$ theory as arising from compactification of  the
heterotic string.  It is well known that the $4D$, $N = 1$ field theory limit
of the
heterotic string consists of a   $4D$, $N = 1$ supergravity multiplet plus
a  $4D$, $N = 1$ tensor multiplet (or linear multiplet) coupled to matter.
Under further  reduction to two dimensions the   supergravity multiplet becomes
a  $2D$, $N
= 2$ supergravity multiplet (plus matter vector multiplets) and the
 tensor multiplet becomes a $2D$, $N = 2$ tensor multiplet which is equivalent
to a  $2D$ vector multiplet.  If we have started with the minimal  version of
$4D$
supergravity,  the reduction  will lead to the $2D, \, U_A (1)$
theory,  while the reduction of the tensor multiplet leads to
a twisted vector multiplet.  Therefore,  a heterotic string compactified
 so that its target space possesses
a realization of $2D$, $N = 2$ supersymmetry naturally leads to the
appearance of the $U_V (1) \otimes U_A (1)$ theory coupled to matter.

The coupling to the  $U_V(1) \otimes U_A(1)$ theory involves two issues. The
first one concerns the existence of  appropriate local measures, the second one
the derivation and form of the projection formulae. There is no difficulty in
generalizing global $D$-terms to local ones and in particular, as we have noted
earlier,
the vielbein determinant is the same as in the $U_A(1)$ theory, $\Hat{E}^{-1} =
E^{-1}$.
Furthermore, in \cite{measures} we  established the existence of
a chiral measure by constructing it in terms of the chiral measure of the
$U_A(1)$
theory, and inferred the existence of a twisted chiral measure.  Equivalently,
the relation between the measures  can be  deduced from \reff{ent} which
allows us
to relate $\Hat{\Delb}^2$ and $\Delb^2$.

The derivation of the chiral  projection formula  \reff{chproj} follows
precisely
 that for the $U_A(1)$
theory. Similarly, the derivation of the twisted chiral projection formula
\reff{twist density}
goes through as in  \cite{measures}, with due regard to the presence
 of the supergravity field strength $\bar{F}$ in the anticommutator $ \{ {\Hat
\nabla}_+ ~,~ {\Hat \nabla}_{\md} \} $. In particular, we find now complete
symmetry between the projection formulae  \reff{chint} and \reff{twist
density}, with the  component auxiliary field $\bar{F}|$
entering in \reff{twist density} in the same manner that the  component
auxiliary field
$\bar{R}| = \bar{B}$ enters in \reff{chint}.

We come to the conclusion that most of the results of the previous sections can
be immediately generalized to the present situation. Obviously, since the
extended
supergravity theory contains more component fields (the components of the
entangled vector multiplet), we may expect these  components to appear
in the final form of results which previously involved just the ordinary
supergravity component fields, namely $B$, the gravitino, and the component
curvature.

\sect{Quantum Considerations}
We now examine at the quantum level the coupling of matter superfields
 to the $U_V(1) \otimes U_A(1)$ version of
supergravity and determine the induced action.  As usual it is sufficient to
compute
the one-loop contribution of chiral
and twisted chiral superfields to the self-energy of the prepotential $H^a$ and
covariantize the result.

  We collect together the expressions that we need for various
quantities in the $U_V(1) \otimes U_A(1)$ supergravity theory at the linearized
level, in terms of  the prepotentials $H^a$ and $S$ \cite{MGMW}.
For the field strengths we have
\bea
\Rb &=& -D_+ \G_- - D_- \G_+   \non \\
\Fb &=& 2iD_+\S_\md +2i D_\md\S_+
\eea
and for the connections,
\bea
\G_- &=& -2D_-(S + \bar{S}) -D_+ \Amp \non \\
\G_+ &=&  2D_+(S + \bar{S}) +D_- \Apm \non \\
\S_+ &=& -2iD_+ \bar{S} - i D_- \Apm \non\\
\S_\md &=& 2iD_\md S + i D_\pd \Ampd ~~.
\eea
To first order the $A$'s are given by
\bea
\Amp = i\Cmpp = -2D_\pd D_- H^\pp &~,~& \Apm = i \Cpmm = -2D_\md D_+ H^\mm
\non \\
\Ampd = i\Cpmp = -2D_\md D_+ H^\pp &~,~& \Apmd = i \Cmpm = -2D_\pd D_- H^\mm
{}~~.
\eea

  Upon substitution we obtain
\bea
\Rb &=& 4D_+ D_- (S + \bar{S}) \non\\
R &=& -4D_\pd D_\md (S + \bar{S}) \non \\
F &=& 4D_\pd D_- [ S + \bar{S} + D_+ D_\pd H^\pp - D_\md D_- H^\mm] \non \\
        &=& 4 D_\pd D_- [ S+ \bar{S} + i\di_\pp H^\pp - i \di_\mm H^\mm] \non\\
\Fb &=& -4 D_+ D_\md [ S + \bar{S} - D_\pd D_+ H^\pp + D_- D_\md H^\mm] \non \\
    &=& -4 D_+D_\md [ S+ \bar{S} - i \di_\pp H^\pp + i \di_\mm H^\mm] ~~.
\label{RF}
\eea

We begin with the kinetic action for {\em covariantly} chiral and twisted
chiral superfields,
$S  =  \int d^6z E^{-1} ( \Phib \Phi - \bar{\calX} \calX) $. (In this section
{\em ordinary} chiral and twisted chiral superfields are denoted by $\phi$ and
$\chi$,
respectively.)  We have $\Phi = e^{iH \cdot \di} \phi$, but a closed expression
for  ${\cal X}$
in terms of $\chi$ is not known.  However in \cite{measures},  we derived an
expression relating a covariantly twisted chiral superfield to an
ordinary one at the linearized level,  which is sufficient for our
 present one-loop calculation,
\bea
\calX &=& [1-2 D_\pd H^\pp D_+ + 2 D_- H^\mm D_\md  + i H^\mm \di_\mm
   - i H^\pp \di_\pp] \chi  \non \\
\bar{\calX} &=& [1+2 D_+ H^\pp D_\pd - 2 D_\md H^\mm D_-  -
            i H^\mm \di_\mm  + i H^\pp \di_\pp] \chib ~~.
\eea
We also expand  the inverse of the superdeterminant to first order in $H^a$
\beq
E^{-1} = 1 - [D_\pd , D_+] H^\pp - [D_\md, D_-] H^\mm ~~.
\eeq

We obtain the action from which we can compute the one-loop contribution,
\bea
S = \int d^6z
&[ &\bar{\phi} \phi -2 H^\pp D_\pd \bar{\phi} D_+ \phi
         -2 H^\mm D_\md \bar{\phi} D_-
 \phi + \cdots ] \non \\
 -&[&  \chib \chi + 2H^\pp D_\pd \chib D_+ \chi
         -2H^\mm D_-\chib D_\md \chi + \cdots]  ~~ .
\eea
For the propagators, we have
\beq
< \phib (z) \phi (z') > = - \frac{D^2 \Db^2}{\pa_\mm \pa_\pp} \d^{(2)}
(z-z') \d^{(4)}(\th_z -\th_{z'}) ~~,
\eeq
and
\beq
< \chib (z) \chi (z') > = - \frac{D_+ D_\md D_- D_\pd}{\pa_\mm \pa_\pp}
    \d^{(2)}(z-z') \d^{(4)}(\th_z -\th_{z'} ) ~~.
\eeq

 We find that the contribution to the $H^a$ self-energy at one-loop is
\bea
-\frac{8}{\pi} \Big[ D_\pd D_- \di _\pp H^\pp \frac{1}{\Box} D_+ D_\md \di _\pp
H^\pp
+ D_\pd D_- \di _\mm H^\mm \frac{1}{\Box} D_+ D_\md \di _\mm H^\mm  \Big] ~~.
\eea
In order to covariantize the result, we add in extra local cross-terms which
give
\beq
-\frac{8}{\pi}[ D_\pd D_- (\di _\pp H^\pp -\di _\mm H^\mm) \frac{1}{\Box}
                           D_+ D_\md (\di _\pp H^\pp -\di _\mm H^\mm) ] ~~,
\eeq
and  using  \reff{RF} we rewrite the result in  covariant form, representing
the induced
action
\beq
 {\cal S}_{ind} =  \frac{1}{4\pi} \int d^2x d^4 \th\Big[ \bar{R}
\frac{1}{\Box}R
- \bar{F} \frac{1}{\Box}F\Big] \non ~~,
\eeq
which is manifestly mirror symmetric.

 As formulated in this paper, the mirror
transformation is a classical discrete transformation defined on rigid (2,2)
supersymmetric off-shell field theories and the coordinates of rigid (2,2)
superspace.  The classical mirror symmetric theories are those that are
invariant
under this discrete transformation. By coupling to the $U_V (1) \otimes
U_A (1)$ supergravity we realize mirror symmetry within the context of a
locally
supersymmetric theory.  The calculation above shows that this discrete
symmetry can be preserved after quantization.

\sect{Concluding remarks}
We have seen that the mirror transformation is an  intrinsic
geometric feature of $2D$, $N \,=\, 2$ superspace theories. It has been
conjectured that this is the origin of mirror manifolds as seen in the (2,2)
Calabi-Yau compactification of the heterotic string, but the precise connection
is unclear, as is the role of duality as an attempt to relate theories  that
are connected by the mirror map operator. Stated in its
simplest form, is there a first order formulation such that eliminating some
of its variables yields (7.28), while a different elimination yields
(7.29)?  A second such question is whether there exists a relation between
the strong coupling limit of (7.28) and the weak coupling limit of (7.29).

We have presented in this paper a number of tools and results concerning
the coupling of $2D$, $N=2$ matter multiplets to supergravity. Although we
have not  studied in detail linear complex multiplets, it is clear that our
results can be easily generalized to them. We would be led to consider
therefore general models of the form
\bea
{\cal S} &=& \int d^2 x \, d^4 \th ~ E^{-1} \,
{\Hat \O} {~~~~~~} \cr
&{~}&  +~ \Big[ ~\int d^2 x \, d^2 \th ~ {\cal E}^{-1} \, [ ~h_{{\rm I
 \, J}} \, (\Del_{\pd} \S^{\rm I} \,)  (\Del_{\md} \S^{\rm J} \,)
{}~+~ U ~ ] ~+~ {\rm {h. \, c.}} ~ \Big] \cr
&{~}& +~ \Big[ ~\int d^2 x \, d \th^+ d \th^{\md} ~ {\Tilde {\cal E
}}^{-1} \, [ ~ k_{i \, j} \, (\Del_+ \Xi^i \,) (\Del_{\md} \Xi^j \,)
{}~+~ {\Hat U} ~ ] ~+~ {\rm {h. \, c.}} ~ \Big]  ~~~~.
\eea
The functions that govern the form of this general action include
the real K\" ahler-like potential ${\Hat \O}$, holomorphic target space
second-rank ({\em not} necessarily skew symmetric) tensors $h_{{\rm
I \, J}}$ and $k_{i \, j}$ and holomorphic target space scalars $P$ and
${\Hat P}$. The K\" ahler-like potential may depend on {\em any}
scalar superfield whose physical degrees of freedom consist of spins
0, 1/2 including $2D$, $N = 2$ vector multiplets.  These
vector multiplets are described in  terms of superspace covariant
derivatives that have a
${\cal G}_c \otimes {{\cal G}'}_{tc}$ structure, where ${\cal G}_c$
acts on chiral and complex linear superfields and ${{\cal G}'}_{tc}$
acts on twisted chiral and twisted complex linear superfields.

The existence
of the holomorphic second-rank tensors $h_{{\rm I \, J}}$ and $k_{i \, j}$
has not been generally recognized previously. The former is restricted
to depend solely on chiral superfields and the latter solely on twisted
chiral supefields. This is the same as for the holomorphic target space
scalars $U$ and ${\Hat U}$, respectively. The off-shell supergravity
fields that appear in the action may be chosen to be either the $U_V (1)$,
$U_A (1)$ (minimal irreducible) or $U_V (1) \otimes U_A (1)$ theories. The
rigid version of this action is expected to provide the most general
off-shell (2,2) compactification of the heterotic string.  To  {\em
every} action of this form there is naturally associated a mirror reflected
theory.  Thus, (2,2) supersymmetric theories
necessarily come in pairs with the mirror map transformation providing the
isomorphism between members of the pairs.  Although we have not investigated
the issue,  it may be possible extend the notion of the
mirror map to act from the space of chiral superfields to
(twisted) complex linear superfields. Finally, we have every reason to believe
that many  of our results can be extended  to $2D$,
$N = 4$ theories.

\noindent {\bf{Acknowledgment} }
M.T.G. and M.E.W.  thank the Physics Departments of Harvard University and the
University
of Maryland for hospitality during the period when this work was done, and
M.E.W.
thanks the Physics Department of Queen's University for hospitality.
S.J.G. thanks Tristan H\"{u}bsch for discussions.

\newpage

\appendix
\section{{\bf Appendix}}
\setcounter{equation}{0}

We restrict ourselves to the minimal $U_A(1)$ theory by setting $\Sigma_A =0$.
In defining components we follow the philosophy and methods described in
{\em Superspace}, \cite{bible}.
  The expressions for the covariant derivatives evaluated at $\th=0$ are:
\bea
\Dela | &=& \di_\a   \non \\
\Del_a | &=& \bD_a + \psi_a^{\a} \Dela| + \psi_a^{\ad} \Delad| \label{Del|}
\non \\
         &=& \bD_a + \psi_a^{\a} \di_{\a} + \psi_a^{\ad} \di_{\ad} ~~.
\eea

In the $U_A(1)$ theory we define $\bD_a$ as the fully covariant gravitational
derivative with a
Lorentz
connection that includes, in addition to the ordinary connection, extra terms
that are bilinear in the gravitini. Specifically, $\bD_a$ is defined to
be
\bea
  \bD_a &=& e_a + {\vf}_a {\cal M} +  V' _a  {\cal Y}'  \non \\
        &=& e_a + \o_a M + \g_a \Mb  \,
\eea
where $\o_a$ is the gravitational plus $U(1)$ component connection, including
gravitini.

 We also introduce the ordinary gravitational
covariant derivative (without $U(1)$ connection or gravitino terms), denoted
${\cal D}_a$
\beq
  {\cal D}_a = e_a + {\vp}_a {\cal M}  ~~~
\eeq
and in particular define $\pa_\pp \equiv e_\pp= e_\pp^m \pa_m$,
 $\pa_\mm \equiv e_\mm= e_\mm^m \pa_m$.

The  component connections are defined by projection to be:
\bea
        \F_A | &=& \vf_A  \non  \\
       \S_A' | &=& V' _A   \non    \\
        \S_A | &=& V_A       \,
\eea
and
\bea
      \O_A | &=& \o_A \non  \\
      \G_A | &=& \g_A ~~~.
\eea
Note that
\beq
\vf_A = \half (\o + \g)_A ~~{\rm and} ~~  V' _A = \sihalf (\o - \g)_A  ~~.
\eeq

 We  also need expressions for objects such as $\Dela \Del_{\b}|$, for
example.  The $\Dela$ component of $\Del_{\b}$ is given by (c.f.
{\em Superspace} sec. 5.6.b)
\beq
\Dela \Del_{\b}| = \half \{\Dela , \Del_{\b}\}|  \label{alphbeta}
\eeq
while the $\Dela$ component of $\Del_b$ is
\bea
\Dela \Del_b | &=& [\Dela , \Del_b]| + \Del_b \Dela | \non \\
               &=& [\Dela , \Del_b]| + \bD_b \Dela | + \psi_b^{\b} \di_{\b}
                   \Dela | + \psi_b^{\bd} \di_{\bd} \Dela | ~~~. \label{abeta}
\eea
{}From \reff{alphbeta} we obtain the following relations
\bea
\Delp \Delp | &=& \half \{\Delp , \Delp \} = 0  \non \\
\Delm \Delm | &=& \half \{\Delm , \Delm \} = 0 \non \\
\Delp \Delpd | &=& \half \{\Delp , \Delpd \} = \sihalf \Delpp \non \\
\Delm \Delmd | &=& \half \{\Delm , \Delmd \} = \sihalf \Delmm \non  \\
\Delp \Delm | &=& \half \{\Delp , \Delm \} = - \half \Rb | \Mb  \non\\
\Delp \Delmd | &=& \half \{\Delp , \Delmd \} = 0  \non\\
\Delm \Delpd | &=& \half \{\Delm , \Delpd \} = 0 ~~,
\eea
and from \reff{abeta}, we get the series of identities that appears below.
\bea
\Delp \Delpp | &=&  [\Delp, \Delpp]| + \Delpp \Delp | \non \\
&=& \bDpp \Delp| + \psi_{\pp}^- \Delm \Delp| + \psi_{\pp}^{\pd} \Delpd \Delp|
     \non \\
&=& \bDpp \Delp| - \half \psi_{\pp}^- \Rb| \Mb + \sihalf \psi_{\pp}^{\pd}
   (\bDpp + \psi_{\pp}^{\a} \Del_{\a} + \psi_{\pp}^{\ad} \Del_{\ad})|   \non\\
\Delm \Delpp | &=&  [\Delm, \Delpp]| + \Delpp \Delm | \non \\
&=& \bDpp \Delm| + \psi_{\pp}^+ \Delp \Delm| + \psi_{\pp}^{\md} \Delmd \Delm |
    + \sihalf \Rb \Delpd | -i(\Delpd \Rb) \Mb | \non \\
&=& \bDpp \Delm| - \half \psi_{\pp}^+ \Rb| \Mb + \sihalf \psi_{\pp}^{\md}
   (\bDmm + \psi_{\mm}^{\a} \Del_{\a} + \psi_{\mm}^{\ad} \Del_{\ad})| +
   \sihalf \Rb \Delpd | -i(\Delpd \Rb) \Mb | \non \\
\Delpd \Delpp |
&=& \bDpp \Delpd| + \half \psi_{\pp}^{\md} R| M + \sihalf \psi_{\pp}^+
   (\bDpp + \psi_{\pp}^{\a} \Del_{\a} + \psi_{\pp}^{\ad} \Del_{\ad})|  \non\\
\Delmd \Delpp |
&=& \bDpp \Delmd| + \half \psi_{\pp}^{\pd} R| M + \sihalf \psi_{\pp}^-
   (\bDmm + \psi_{\mm}^{\a} \Del_{\a} + \psi_{\mm}^{\ad} \Del_{\ad})| -
   \sihalf R \Delp | +i(\Delp R) M | \non \\
\Delp \Delmm |
&=& \bDmm \Delp| - \half \psi_{\mm}^- \Rb| \Mb + \sihalf \psi_{\mm}^{\pd}
   (\bDpp + \psi_{\pp}^{\a} \Del_{\a} + \psi_{\pp}^{\ad} \Del_{\ad})| -
   \sihalf \Rb \Delmd | -i(\Delmd \Rb) \Mb | \non  \\
\Delm \Delmm |
&=& \bDmm \Delm| - \half \psi_{\mm}^+ \Rb| \Mb + \sihalf \psi_{\mm}^{\md}
   (\bDmm + \psi_{\mm}^{\a} \Del_{\a} + \psi_{\mm}^{\ad} \Del_{\ad})|  \non\\
\Delpd \Delmm |
&=& \bDmm \Delpd| + \half \psi_{\mm}^{\md} R| M + \sihalf \psi_{\mm}^+
   (\bDpp + \psi_{\pp}^{\a} \Del_{\a} + \psi_{\pp}^{\ad} \Del_{\ad})| +
   \sihalf R \Delm | +i(\Delm R) M | \non \\
\Delmd \Delmm |
&=& \bDmm \Delmd| + \half \psi_{\mm}^{\pd} R| M + \sihalf \psi_{\mm}^-
   (\bDmm + \psi_{\mm}^{\a} \Del_{\a} + \psi_{\mm}^{\ad} \Del_{\ad})|
\eea
where we have used the commutation relations
\bea
{[}\Delp, \Delpp] &=& 0 ~~~, ~~~[\Delm, \Delmm ] = 0   \non\\
{[}\Delpd, \Delpp ] &=& 0 ~~~, ~~~[\Delmd, \Delmm ] = 0  \non \\
{[}\Delp, \Delmm ] &=& - \sihalf \Rb \Delmd  -i(\Delmd \Rb) \Mb  \non\\
{[}\Delpd, \Delmm ] &=&  \sihalf R \Delm  +i(\Delm R) M  \non\\
{[}\Delm, \Delpp ] &=&  \sihalf \Rb \Delpd  -i(\Delpd \Rb) \Mb  \non\\
{[}\Delmd, \Delpp ] &=&  -\sihalf R \Delp  +i(\Delp R) M
\eea
and also
\bea
[\Delpp, \Delmm ] &=&  \half (\Delp R) \Delm +\half (\Delm R) \Delp
        - \half (\Delpd \Rb) \Delmd - \half (\Delmd \Rb) \Delpd \non \\
&& -  \half R \Rb \Mb - \half R \Rb M + (\Delsq R) M -(\Delbsq \Rb) \Mb
\label{DppDmm} ~~.
\eea
We also find the $\Delpp$ component of $\Delmm$ to be:
\bea
\Delpp \Delmm | &=& (\bDpp + \psi_{\pp}^{\a} \Del_{\a} +
                               \psi_{\pp}^{\ad} \Del_{\ad}) \Delmm| \non  \\
     &=& \Delpp| \Delmm| + \psi_{\pp}^{\a} \Del_{\a} \Delmm| +
                               \psi_{\pp}^{\ad} \Del_{\ad} \Delmm| ~~ .
\eea
Using this we can show that
\beq
[\Delpp, \Delmm] | = [\Delpp |, \Delmm |] +  \psi_{\pp}^{\a} \Del_{\a} \Delmm|
   + \psi_{\pp}^{\ad} \Del_{\ad} \Delmm| - \psi_{\mm}^{\a} \Del_{\a} \Delpp |
   - \psi_{\mm}^{\ad} \Del_{\ad} \Delpp|  ~,
\eeq
where the first term can be expanded as
\bea
[\Delpp |, \Delmm |] &=& [\bDpp +  \psi_{\pp}^{\a} \di_{\a} + \psi_{\pp}^{\ad}
    \di_{\ad}, \bDmm + \psi_\mm^{\b} \di_{\b} + \psi_\mm^{\bd} \di_{\bd}]
    \non \\
&=& [\bDpp, \bDmm] + \bD_{[\pp} (\psi_{\mm]}^{\a} \di_{\a}) +
      \bD_{[\pp} (\psi_{\mm]}^{\ad} \di_{\ad})  \non\\
&=& [\bDpp, \bDmm] + (\bD_{[\pp} \psi_{\mm]}^{\a}) \di_{\a} +
      (\bD_{[\pp} \psi_{\mm]}^{\ad}) \di_{\ad} \non \\
&& + \half \o_{[\pp} (\psi_{\mm]}^+ \di_+ - \psi_{\mm]}^- \di_-)
    + \half \g_{[\pp} (\psi_{\mm]}^\pd \di_\pd - \psi_{\mm]}^\md \di_\md)~~.
\eea
Substituting this into $[\Delpp, \Delmm] |$, we get
\bea
\lefteqn{[\Delpp, \Delmm] | = [\bDpp, \bDmm] + \bD_{[\pp} (\psi_{\mm]}^{\a}
     \di_{\a})
    +  \bD_{[\pp} (\psi_{\mm]}^{\ad} \di_{\ad}) } \non \\
 && \hskip 6em + \psi_{\pp}^{\a} \Del_{\a} \Delmm|
    + \psi_{\pp}^{\ad} \Del_{\ad} \Delmm| - \psi_\mm^{\a} \Del_{\a} \Delpp |
     - \psi_\mm^{\ad} \Del_{\ad} \Delpp| \non \\
 &=& [\bDpp, \bDmm] + \bD_{[\pp} (\psi_{\mm]}^{\a} \di_{\a}) +
      \bD_{[\pp} (\psi_{\mm]}^{\ad} \di_{\ad}) \non \\
 && +  \psi_{\pp}^+ [\Delmm \di_+ + \sihalf \psi_\mm^{\pd}(\bD_{\pp} +
      \psi_{\pp}^{\a} \di_{\a} + \psi_{\pp}^{\ad} \di_{\ad}) - \half
      \psi_\mm^- \Rb \Mb - \sihalf \Rb \di_{\md} - i (\Delmd \Rb) \Mb] \non \\
 && + \psi_{\pp}^- [\bDmm \di_- - \half \psi_\mm^+ \Rb \Mb + \sihalf
\psi_\mm^{\md}
   (\bDmm + \psi_\mm^{\a} \di_{\a} + \psi_\mm^{\ad} \di_{\ad})] \non \\
  && + {\rm~6~other~similar~terms~}. \label{DppDmm|}
\eea
   If we compare this now with \reff{DppDmm} (which
is true in general, not just at $\th=0$), we can obtain expressions for the
various derivatives of $R$ and $\Rb$, evaluated at $\th=0$.  By comparing
the coefficients of $\Delp | = \di_+$ on both sides, for example, we get:
\beq
 \half \Delm R | = \bD_{[\pp} \psi_{\mm]}^+ + \sihalf \psi_\mm^{\md} R| +
 i \psi_{\pp}^- \psi_\mm^{\md} \psi_\mm^+ + i \psi_{\pp}^{\pd} \psi_\mm^+
\psi_{\pp}^+
 + i \psi_{\pp}^{\md} \psi_\mm^- \psi_\mm^+ ~~ .
\eeq
In the same fashion we obtain:
\bea
 \half \Delp R | &=& \bD_{[\pp} \psi_{\mm]}^- + \sihalf \psi_{\pp}^{\pd} R| +
 i \psi_{\pp}^+ \psi_\mm^{\pd} \psi_{\pp}^- + i \psi_{\pp}^- \psi_\mm^{\md}
\psi_\mm^-
 + i \psi_{\pp}^{\pd} \psi_\mm^+ \psi_{\pp}^-   \non\\
- \half \Delpd \Rb | &=& \bD_{[\pp} \psi_{\mm]}^{\md} - \sihalf \psi_{\pp}^+
\Rb|
+
i \psi_{\pp}^+ \psi_\mm^{\pd} \psi_{\pp}^{\md} + i \psi_{\pp}^{\md} \psi_\mm^+
\psi_{\pp}^{\md}
 + i \psi_{\pp}^{\md} \psi_\mm^- \psi_{\mm}^{\md}  \non \\
- \half \Delmd \Rb | &=& \bD_{[\pp} \psi_{\mm]}^{\pd} - \sihalf \psi_\mm^- \Rb|
+
 i \psi_{\pp}^+ \psi_\mm^{\pd} \psi_{\pp}^{\pd} + i \psi_{\pp}^{\md} \psi_\mm^-
\psi_\mm^{\pd}
 + i \psi_{\pp}^- \psi_\mm^{\md} \psi_\mm^{\pd}  \non \\
(-\half R \Rb + \Delsq R)| &=& [\bDpp, \bDmm]_M - i \psi_\mm^{\md} \Delp R|
 +i \psi_{\pp}^{\pd} \Delm R| - \psi_\mm^{\md} \psi_{\pp}^{\pd} R|
 - \psi_\mm^{\pd} \psi_{\pp}^{\md} R|  \non \\
(-\half R \Rb - \Delbsq \Rb)| &=& [\bDpp, \bDmm]_{\Mb} - i \psi_{\pp}^+ \Delmd
\Rb|
+i \psi_\mm^- \Delpd \Rb| - \psi_{\pp}^+ \psi_\mm^- \Rb|  + \psi_\mm^+
\psi_{\pp}^-
\Rb|  \non  \\
\eea

   We can get explicit information about the connections from looking at the
commutator of two $\bD$'s.  In particular
\bea
{[}\bDpp, \bDmm] &=& [e_\pp, e_\mm] + e_{[\pp} \o_{\mm]} M + e_{[\pp} \g_{\mm]}
\Mb
  \non \\
&&- \half \o_{\{\mm} e_{\pp\}} - \half \g_{\{\pp} e_{\mm\}} - \o_\pp \o_\mm M
   \non \\
&&+ \g_\pp \g_\mm \Mb - \half \g_{\{\pp} \o_{\mm \}} M -
    \half \o_{\{\mm} \g_{\pp \}} \Mb ~~~.   \label{dd}
\eea
The anholonomy coefficients are defined as usual by $[e_\pp, e_\mm] =
{C_{\pp \mm}}^a e_a$. Denoting the ``component" torsion by ${t_{\pp \mm}}^a$
where
\beq
[\bDpp, \bDmm] = {t_{\pp \mm}}^a \bD_a + r_{\pp \mm} M + \bar{r}_{\pp \mm} \Mb
{}~,
 \label{dd2}
\eeq
and comparing \reff{dd2} with \reff{dd}, we find the torsions
\bea
{t_{\pp \mm}}^\pp &=& {C_{\pp \mm}}^\pp - \half(\o + \g)_\mm \label{t} \non\\
{t_{\pp \mm}}^\mm &=& {C_{\pp \mm}}^\mm - \half(\o + \g)_\pp  ~~.
\eea
and the curvatures
\bea
r_{\pp \mm} &=& e_{[\pp} \o_{\mm]} - {C_{\pp \mm}}^\pp \o_\pp - {C_{\pp \mm}}
  ^\mm \o_\mm  \non \\
{\bar{r}}_{\pp \mm} &=& e_{[\pp} \g_{\mm]} - {C_{\pp \mm}}^\pp \g_\pp
   - {C_{\pp \mm}}^\mm \g_\mm ~~.
\eea
The full torsions are again defined in the standard way, $[\Del_A, \Del_B \}
= {T_{AB}}^C \Del_C + R_{AB}M + \Rb_{AB}\Mb$.  Therefore, from \reff{DppDmm|}
at $\th = 0$ and using \reff{Del|}, we get the full torsions to be
\bea
{T_{\pp \mm}}^\pp &=& {t_{\pp \mm}}^\pp + i (\psi_\pp^+ \psi_\mm^\pd +
       \psi_\pp^\pd \psi_\mm^+)  \non\\
{T_{\pp \mm}}^\mm &=& {t_{\pp \mm}}^\mm + i (\psi_\pp^- \psi_\mm^\md +
       \psi_\pp^\md \psi_\mm^-) \label{Tmm}  \non\\
{T_{\pp \mm}}^+ &=& \bD_{[\pp} \psi_{\mm]}^+ +\sihalf \psi_\mm^\md R|
      + i \psi_\pp^- \psi_\mm^\md \psi_\mm^+ + i \psi_\pp^\pd \psi_\mm^+
      \psi_\pp^+ + i \psi_\pp^\md \psi_\mm^- \psi_\mm^+  \non \non \\
{T_{\pp \mm}}^- &=& \bD_{[\pp} \psi_{\mm]}^- +\sihalf \psi_\pp^\pd R|
      + i \psi_\pp^+ \psi_\mm^\pd \psi_\pp^- + i \psi_\pp^- \psi_\mm^\md
      \psi_\mm^- + i \psi_\pp^\pd \psi_\mm^+ \psi_\pp^-  ~~.
\eea
It is obvious from \reff{DppDmm} that ${T_{\pp \mm}}^a = 0$.  Combining this
with \reff{t} and \reff{Tmm}, we obtain
\bea
\vf_\pp &=& \half (\o + \g)_\pp \non \\
&=& {C_{\pp \mm}}^\mm + i (\psi_\pp^- \psi_\mm^\md + \psi_\pp^\md \psi_\mm^-)
\non \\
&=& \vp_\pp + i (\psi_\pp^- \psi_\mm^\md + \psi_\pp^\md \psi_\mm^-)
\label{conn1}  \non\\
\vf_\mm &=& \half (\o + \g)_\mm \non \\
&=& {C_{\pp \mm}}^\pp + i (\psi_\pp^+ \psi_\mm^\pd + \psi_\pp^\pd \psi_\mm^+)
\non \\
&=& \vp_\mm + i (\psi_\pp^+ \psi_\mm^\pd + \psi_\pp^\pd \psi_\mm^+)
\label{conn}
\eea
for the full component Lorentz connection, written in terms of the
ordinary component gravitational connection, ${\vp}_a$, plus the gravitini
terms
mentioned previously.

\newpage

\end{document}